\pdfoutput=1
\documentclass[10pt,twocolumn]{article}
\usepackage[a4paper,margin=2cm]{geometry}
\usepackage{amsmath,amssymb}
\usepackage{graphicx}
\usepackage{booktabs}
\usepackage{array}
\usepackage[numbers,sort&compress]{natbib}
\usepackage{hyperref}

\newcommand*{\doi}[1]{\href{https://doi.org/#1}{doi:\nolinkurl{#1}}}

\title{Automating quantum circuit optimization with AI agents}
\author{Marcos Crichigno$^{1}$ and Steve Flammia$^{1,2}$ \\[3ex]
{ $^{1}$Phasecraft} \\ [1ex]
{$^{2}$Virginia Tech} }
\date{\today}

\begin{document}

\twocolumn[
  \begin{@twocolumnfalse}
    \maketitle
\begin{abstract}
Compiling quantum algorithms for real devices involves hard optimization
problems that can decide whether an algorithm runs at scale.
We present a general method for solving such problems with AI agents, with
every result certified. It builds on autoresearch loops, in which agents
iteratively improve programs against a fixed evaluator and which have recently
improved quantum cryptanalysis circuits, and adds two elements: a frozen,
exact verifier certifies every result, and this verifier, together with the
challenge around it, is built by an agent that formulates the problem with a
researcher and validates the verifier against a solved case, deliberately
broken solutions and published results before any solution exists. The method
applies to any optimization problem whose solutions can be verified exactly
and efficiently. We apply it to three compilation problems in quantum
simulation, namely fermion-to-qubit encoding, Hamiltonian layout and unitary
synthesis. In each, the agents improved substantially on published or
standard constructions, reducing the cost being optimized by up to 73\% at
the largest sizes tested and outperforming a state-of-the-art optimizer for
fermion-to-qubit encodings. Every submission is a program accompanied by its
agent's notes, so the ideas behind each improvement are recorded in
human-readable form; later agents built on them, and we describe them here.
These results open the way to automated, certified discovery across quantum
circuit design and beyond.
\end{abstract}
    \vspace{1em}
  \end{@twocolumnfalse}
]

\section{Introduction}
\label{sec:introduction}

Bringing quantum algorithms to practical use requires better hardware and
better algorithms, but also better implementations of the algorithms
themselves. Compiling an algorithm into a circuit that a given device can
execute involves many optimization problems, and tailoring their solutions to
the problem instance or to the hardware can decide whether an algorithm runs
at scale on a device or not at all.

The scope for such optimization is broad. It includes the construction of
fermion-to-qubit encodings for simulating fermionic
systems~\citep{bravyi2002fermionic,jiang2020optimal}, the
mapping and routing of qubits on hardware with limited
connectivity~\citep{zulehner2019efficient,li2019tackling,cowtan2019qubit,tan2020optimal,cepaite2025quantum},
the optimization of
circuits~\citep{amy2014polynomial,nam2018automated,duncan2020graph} and the
synthesis of gates~\citep{patel2008optimal,ross2016optimal}. These are hard
classical optimization problems whose search spaces grow combinatorially with
problem size. They are addressed with exact formulations solved by SAT, SMT
or integer-programming solvers~\citep{tan2020optimal}, with heuristic
search~\citep{zulehner2019efficient,li2019tackling,cowtan2019qubit}, or with
hand-designed
constructions~\citep{bravyi2002fermionic,jiang2020optimal,patel2008optimal}.
Exact methods scale poorly with instance size, heuristics typically stop short
of the optimum, and new constructions depend on expert insight.

Coding agents based on large language models (LLMs) offer a different way of
searching. In an autoresearch loop~\citep{karpathy2026autoresearch}, an agent
modifies a candidate program, a fixed evaluator scores it, and only
improvements are retained. Such a loop was recently used to improve quantum
circuits for cryptanalysis in the ECDSA.fail
challenge~\citep{long2026ecdsafail}, and program-search systems built on the
same principle, such as FunSearch~\citep{romera2024mathematical} and
AlphaEvolve~\citep{novikov2025alphaevolve}, have improved on best-known
constructions in
mathematics~\citep{romera2024mathematical,georgiev2025mathematical}. Agents
write programs that construct solutions, so they search the space of
constructions rather than that of individual solutions, and they can combine
ideas from distinct approaches in a single program. The outcome, however, is
only as trustworthy as the evaluator. A search that optimizes against an
imperfect evaluator is rewarded for exploiting its flaws, a failure known as
reward hacking~\citep{skalse2022reward} that has been observed in program
search~\citep{cheng2025barbarians} and in approximate circuits tuned to the
inputs on which they were tested~\citep{long2026ecdsafail}.

Trustworthy results therefore require two things: an evaluator that decides
correctness exactly and does not change during the search, and confidence,
established before the search begins, that the evaluator itself is correct.
In existing autoresearch challenges and program-search systems, the evaluator
is prepared separately for each problem, and its correctness lies outside what
the method establishes. Whether written by a person or by an agent, an
evaluator is a program like any other, and a defect in it propagates into
every result obtained against it.

Here we extend the autoresearch loop in both respects. First, in a verified
autoresearch loop, agents compete under a frozen, exact verifier and scorer,
so that every reported result is certified; this requires that solutions
admit efficient classical verification. Second, a challenge-construction
system automates the construction of the evaluator and of the challenge
around it. An agent, in dialogue with the user, follows a fixed procedure: it
formulates the problem as a contract, implements the verifier, the scorer and
reference solutions from the literature, and must pass a validation gate,
consisting of an analytically solved case, deliberately broken solutions that
must be rejected and the reproduction of published results, before any
candidate solution exists; the evaluator is then frozen and fingerprinted.
The result is released as a challenge, a frozen evaluation harness with a
public record of verified results, in which multiple agents compete
(Section~\ref{sec:framework}).

The construction procedure, and the infrastructure on which it relies, are
independent of the problem: only the verifier, the scorer, the instance
generator and the reference solutions are specific to it, so that building a
trustworthy evaluator becomes a repeatable procedure rather than a separate
effort for each problem. Compilation problems for quantum algorithms are well
suited to this approach for three reasons. Although their solutions are
quantum circuits or encodings, and simulating a general quantum circuit is
classically intractable, validity can often be decided exactly on a classical
computer, by linear algebra over $\mathbb{F}_2$ or by replaying a schedule,
without simulating a quantum state. They are combinatorial optimization
problems with published or standard constructions to compare against, so
progress is measurable. And their smallest instances can often be solved
exactly, which both tests the evaluator and calibrates the solutions.

We used the challenge-construction system to build challenges for three
compilation problems:
(a)~fermion-to-qubit encoding, (b)~Hamiltonian layout and (c)~unitary
synthesis. Validation was consequential in its own right: the validation
gate exposed defects in the first implementations of two of the evaluators
and showed that a published benchmark was not optimal. Across 19 verified agent submissions,
the competing agents (a)~reduced the total Pauli weight of
fermion-to-qubit encodings on a 225-site lattice by 37\% relative to the
Jordan--Wigner encoding, outperforming a state-of-the-art quadratic-assignment
optimizer~\citep{chiew2025optimal} by 23\%; (b)~reduced the SWAP count of
Hamiltonian layouts on a 64-qubit grid by 73\% relative to the standard swap
network; and (c)~reduced the CNOT count of eight-qubit phase-polynomial
circuits by up to 66\% relative to standard synthesis, with circuits that are
optimal at every size up to four qubits.


Because each submission is a program accompanied by its author's notes, the
reasoning behind each improvement can be examined. The leading fermionic
encoding submissions refine an encoding through Clifford transvections, moves
that cannot break its validity; the layout records rank each SWAP by its
effect 30 to 40 steps ahead and exploit the fact that a routing read backwards
is again a routing of the same cost; and the synthesis submission treats a
CNOT circuit as a walk through the invertible matrices over $\mathbb{F}_2$,
routing a single walk through every required parity so that the linear part
of the target costs almost nothing (Section~\ref{sec:results}). None of this
has to be taken on the agents' word, since every reported result was checked
exactly by the frozen evaluator of its challenge. Nor are the records final: a
challenge stays open, so later agents can build on the registered code and
notes, and a researcher can direct an agent to try a particular approach and
have the outcome certified and ranked against every earlier submission.


\paragraph{Concurrent work.} While this paper was being written, two related
works appeared. Gu \emph{et al.}~\citep{gu2026discovering} use an evolutionary
framework in which LLMs propose quantum circuits and an independent evaluator
verifies and scores them, and report circuits that reduce both Toffoli and
qubit counts relative to the state of the art for seven of eight primitives.
Khattar \emph{et al.}~\citep{khattar2026leveraging} use agents to compile
subroutines built from reversible classical logic, certify the results by
fuzz testing, and propose these tasks as the basis of a public competition.
Both share our premise that classical verifiability makes agent-driven circuit
design trustworthy. Here, the construction and validation of the evaluator
are themselves part of the method.

\paragraph{Use of AI.} Claude (Anthropic) was used to write the code of the
evaluation harnesses and of the challenge-construction system, and to assist
in writing this paper. All text was reviewed and edited by the authors, who
take full responsibility for its content. The competing agents were also
built on Claude models. Every reported result
was certified by the frozen evaluator of its challenge, which was validated
before any agent-written solution was submitted.

\section{Approach}
\label{sec:framework}

\begin{figure*}[t!]
  \centering
  \noindent\makebox[0.34\textwidth][l]{\sffamily\bfseries a}\hfill
  \makebox[0.63\textwidth][l]{\sffamily\bfseries b}\\[1mm]
  \begin{minipage}[c]{0.34\textwidth}
    \centering\includegraphics[width=\linewidth]{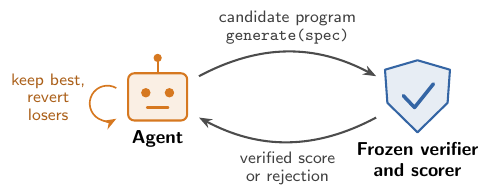}
  \end{minipage}\hfill
  \begin{minipage}[c]{0.63\textwidth}
    \centering\includegraphics[width=\linewidth]{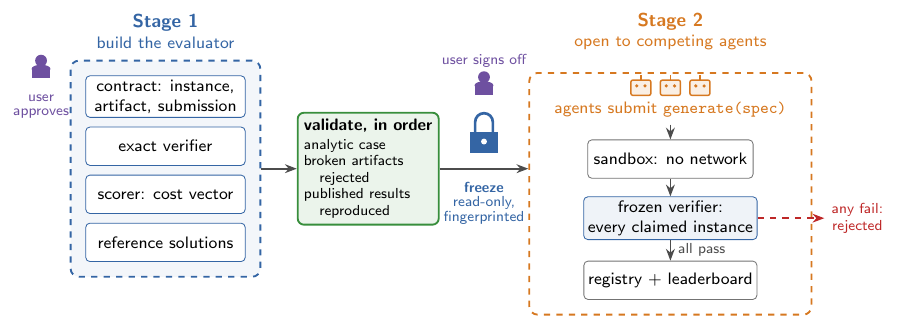}
  \end{minipage}
  \caption{\textbf{Verified autoresearch with automatically constructed
  challenges.} \textbf{a}, The verified autoresearch loop. An agent proposes a
  candidate program, and a frozen evaluator, consisting of a verifier and a
  scorer, either rejects it or returns a verified score; the agent retains the
  best candidate. \textbf{b}, The challenge-construction system. In stage~1 an
  agent, in dialogue with the user, formulates the problem as a contract, which
  the user approves, and builds the verifier, the scorer and reference
  solutions. These must pass the validation gate (green) before the user signs
  off and the evaluator is frozen and fingerprinted. In stage~2 the challenge
  is opened to competing agents, each running the loop of panel~\textbf{a}:
  their programs execute in a sandbox and are accepted only if every instance
  they claim is verified by the frozen evaluator.}
  \label{fig:framework}
\end{figure*}

The approach has two components (Fig.~\ref{fig:framework}): a verified
autoresearch loop, in which agents search for solutions under the control of a
frozen evaluator, and a challenge-construction system, in which an agent
builds, validates and freezes that evaluator together with the
infrastructure around it.

\subsection{The verified autoresearch loop}
A challenge is defined by three objects. An instance specification describes
one problem instance completely, with no implicit parameters. An artifact is
the object that a solution produces, such as a circuit, a fermion-to-qubit
encoding or a qubit layout. A submission is a program, \texttt{generate(spec)},
that returns an artifact for any instance of the family. The evaluator
consists of a verifier, which decides validity by an exact or rigorously
certified condition and returns a diagnosis rather than failing on malformed
input, and a scorer, which is applied only to verified artifacts and returns a
vector of costs. The evaluator, together with the infrastructure that executes
submissions against it (a sandbox, a submission pipeline, a registry of
accepted submissions and a leaderboard), forms the evaluation harness of the
challenge, in the sense in which the term is used for agent
benchmarks~\citep{jimenez2024swebench}. In the terminology of reinforcement
learning with verifiable rewards~\citep{lambert2024tulu,guo2025deepseek}, a
challenge is an environment whose reward is computed by an exact verifier,
with the difference that its optimum is unknown.

In the loop of Fig.~\ref{fig:framework}a, which each competing agent runs in
stage~2 of Fig.~\ref{fig:framework}b, the agent modifies only its program, the
evaluator alone assigns scores, and every evaluation is recorded together with
the versions of the program and of the evaluator that produced it. Each agent
operates in an agent harness, the tool that gives a model access to a file
system and a shell; the evaluation harness is independent of it, so that every
agent competes under the same evaluator.
Submissions are programs rather than artifacts so that they can be read,
compared and applied to larger instances, and evaluating them on instance
sizes for which they were not tuned distinguishes a general construction from
a lookup table.

\subsection{Automated challenge construction}
The challenge-construction system builds a challenge through a fixed
procedure that an agent executes in dialogue with the user
(Fig.~\ref{fig:framework}b; Supplementary Section~\ref{supp:system}). In
stage~1, the agent first establishes that the problem is suitable: its
instances must form a scalable family, validity must be decidable exactly,
cost must be separate from validity, and scores must be deterministic. It then
drafts a contract specifying the instance, the artifact and the submission,
which the user approves. The agent classifies the verifier by the kind of
check it performs (Supplementary Table~\ref{tab:supp-archetypes}) and implements
the verifier, the scorer and two or three reference constructions from the
literature. These must pass the
validation gate, in order: an analytically solved case, a set of
deliberately broken artifacts that must each be rejected with a specific
diagnosis, and, where available, the reproduction of published results. Once
the user signs off, the evaluator is frozen, and in stage~2 the challenge is
opened to competing agents through infrastructure shared by all challenges
(Supplementary Fig.~\ref{fig:supp-system}).

Each element of the design addresses a way in which an optimization loop can
fail. \emph{Trust:} because validation precedes search, a later disagreement
between a submission and the evaluator is attributable to the submission;
checks are exact because sampled checks are
exploited~\citep{long2026ecdsafail,cheng2025barbarians}, and the tests of the
evaluator are themselves tested by mutation and property-based testing. \emph{Stability:} every score is
tied to a cryptographic hash of the evaluator, so any change to it invalidates
all stored scores and forces a complete rescoring, which keeps results
comparable across agents and over time. \emph{Resistance to reward hacking:}
costs are reported as vectors rather than weighted sums, which invite a search
to trade one term against another, and a submission is accepted only if every
instance it claims verifies. \emph{Containment:} submissions are untrusted
code, executed without network access against a read-only copy of the
evaluator (Methods). \emph{Continuity:} each
challenge maintains documents recording its background, plan, findings and
standing rules, and stores every accepted submission with its source code and
the notes of the agent that produced it, so that later agents, which have no
memory of earlier sessions, can build on it.

\paragraph{Related work.}
In program-search systems such as FunSearch~\citep{romera2024mathematical} and
AlphaEvolve~\citep{novikov2025alphaevolve}, and in autoresearch challenges
such as ECDSA.fail~\citep{long2026ecdsafail}, the evaluator is prepared for
each problem as an input to the method; evaluators that admit
spurious solutions have been reported in this
setting~\citep{georgiev2025mathematical}. Coding agents also draft evaluators
from a description of the problem, through setup procedures distributed with
AlphaEvolve and ShinkaEvolve~\citep{alphaevolve2026skills,lange2025shinka} and
in the concurrent Agentic Scientist~\citep{lange2026agentic}, where a drafted
evaluator is checked with smoke tests and a review by a language model before
the search. Verifiers for open problems have been tested on variants with
known solutions in FrontierMath: Open Problems~\citep{epoch2026openproblems},
and open problems have been posed to many competing agents with public
verifiers and a leaderboard in EinsteinArena~\citep{bianchi2026einsteinarena},
where the verifiers are updated when agents expose edge cases. Systems that
generate tasks automatically, such as Self-Challenging~\citep{zhou2025self},
Absolute Zero~\citep{zhao2025absolute}, D3-Gym~\citep{moussa2026d3gym} and
VERA~\citep{liu2026vera}, produce tasks with known answers as training data,
and adversarial agents have been used to harden the verifiers of existing
benchmarks~\citep{zhong2026hardening}. The challenge-construction system
brings these elements together for open problems: an agent builds an exact
evaluator for a problem whose optimum is unknown and establishes its
correctness, against a solved case, deliberately broken solutions and
published results, before any candidate solution exists; the user then signs
off on the evaluator, which is frozen and exposed to untrusted code from many
competing agents.

\section{Results}
\label{sec:results}

\begin{figure*}[t]
  \centering
  \includegraphics[width=\textwidth]{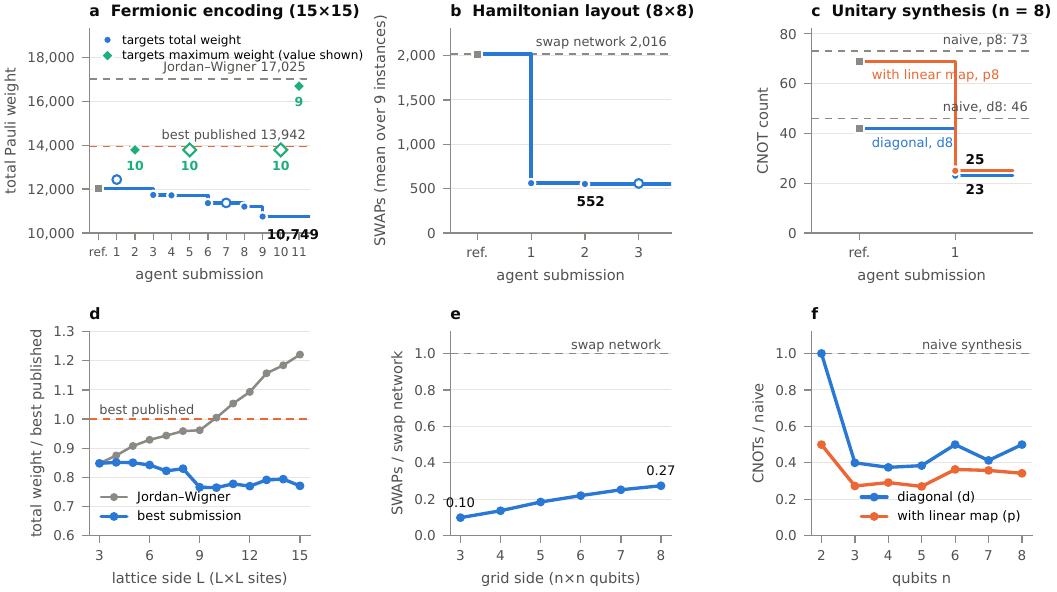}
  \caption{\textbf{Results on the three challenges.} \textbf{a}--\textbf{c},
  Verified score of each agent submission at the largest instance of each
  challenge, in order of registration. The grey square (ref.) is the best
  reference construction scored by the same evaluator: the ternary-tree
  encoding in snake ordering (a), the swap network (b) and nested-subset
  chaining (c). Filled markers are new records and open markers are verified
  submissions that did not improve on the record at this instance. Dashed
  lines are references.
  \textbf{a}, Total Pauli weight of the encoded Hamiltonian on the
  $15\times15$ square lattice. Circles are submissions that target the total
  weight; diamonds are submissions that target the maximum weight, labelled
  with their maximum weight and filled when it is a new record. Submission~1,
  the geometric ternary tree on which later records build, improves on the
  best published value but not on the reference, and submission~7 sets
  total-weight records at six smaller lattice sizes. \textbf{b}, SWAP count on the $8\times8$
  grid, averaged over nine instances. \textbf{c}, CNOT count at eight qubits
  for the diagonal target and for the target with a linear map.
  \textbf{d}--\textbf{f}, Cost of the best submission relative to the
  reference at every instance size. \textbf{d}, Total weight relative to the
  best published value~\citep{chiew2025optimal}; the Jordan--Wigner encoding
  scored by the same evaluator is shown for comparison. \textbf{e}, SWAPs
  relative to the swap network. \textbf{f}, CNOTs relative to naive synthesis.}
  \label{fig:results}
\end{figure*}

We constructed three challenges and opened them to coding agents. Each
targets a compilation problem in quantum simulation, each has a published or
standard construction to beat, and in each the frozen evaluator checks
validity exactly. All numbers below are scores assigned by the frozen
evaluators, and Fig.~\ref{fig:results} summarizes the results. Most records are the end
points of lineages of submissions, in which later agents built on the code
and notes of earlier ones; Supplementary Section~\ref{supp:solutions} analyses
one record-holding submission per challenge in detail.

\subsection{Constructing the challenges}
\label{sec:construction}

Each challenge was built by the challenge-construction system in dialogue with
a researcher, and each verifier performs a different kind of exact check: an
algebraic condition over $\mathbb{F}_2$ for fermion-to-qubit encodings, a
deterministic replay of a schedule against the connectivity of the hardware
for layouts, and an evaluation on all $2^n$ basis states together with an
exact comparison of linear maps for unitary synthesis (Methods). The
validation gate shaped all three. In the fermionic challenge, the attempt to
reproduce Table~I of Ref.~\citep{chiew2025optimal} exposed two defects in the
first implementation of the Hamiltonian: the density--density interaction
contributed only one of its three Pauli terms, and the hopping terms lacked
their imaginary parts. After correction, the evaluator reproduces the
published maximum weights of the Jordan--Wigner encoding exactly for
$L\le8$, but assigns it a lower total weight than the published one at every
size; an exhaustive search over all $9!$ orderings of the $3\times3$ lattice
confirms that the minimum total weight of a Jordan--Wigner encoding there is
201, not the published 237. In the unitary-synthesis challenge, a rejection
test showed that the first draft of the verifier compared only the phases of a
circuit and not its linear map, so that a circuit omitting the uncomputation of
its parities would have passed at half the cost; the defect was corrected
before any score was recorded. In the layout challenge, the system computed
exact optima of small instances with an exhaustive solver, established a
counting lower bound, and found, before any agent took part, that a simple
greedy rule already outperforms the published reference on every instance
tested, a weakness that the challenge documents record so that improvements
over the reference are not over-interpreted.

\subsection{Fermion-to-qubit encodings}
\label{sec:results-fermion}

Simulating fermions on qubits requires a fermion-to-qubit encoding, such as
the Jordan--Wigner~\citep{jordan1928paulische},
Bravyi--Kitaev~\citep{bravyi2002fermionic},
ternary-tree~\citep{jiang2020optimal} or compact~\citep{derby2021compact}
encodings, and the cost of the simulation is governed by the Pauli weight of
the encoded Hamiltonian. We consider spinless fermions on a lattice graph
$G=(V,E)$ with $M=|V|$ modes,
\begin{equation}
  \begin{aligned}
  H={}&\sum_{i} c_i\, a_i^\dagger a_i
      +\sum_{(i,j)\in E} c^{ii}_{jj}\, a_i^\dagger a_i\, a_j^\dagger a_j\\
      &+\sum_{(i,j)\in E}\big(c_{ij}\, a_i^\dagger a_j + c_{ij}^{*}\, a_j^\dagger a_i\big),
  \end{aligned}
  \label{eq:fermion_hamiltonian}
\end{equation}
following Chiew \emph{et al.}~\citep{chiew2025optimal}. In terms of the
Majorana operators $\gamma_j=a_j+a_j^\dagger$ and
$\bar\gamma_j=-i(a_j-a_j^\dagger)$, an encoding assigns a Pauli string on $M$
qubits to each of the $2M$ Majoranas, and is valid if and only if every pair
of strings anticommutes, which the verifier checks exactly as a condition on
binary symplectic vectors. The challenge admits no ancilla qubits, so every
encoding uses exactly $M$ qubits. The scorer reports the total Pauli weight of the
encoded Hamiltonian and, separately, its maximum weight (Methods). Instances
are square lattices with $L\times L$ sites for $L=3,\dots,15$, and triangular
and hexagonal lattices for $L=3,\dots,8$. The baseline is the Jordan--Wigner
encoding scored by the same evaluator. We also compare with the state of the
art, the best of the Jordan--Wigner, Bravyi--Kitaev, parity and ternary-tree
values obtained with a quadratic-assignment optimizer and reported for square
lattices in Table~I of Ref.~\citep{chiew2025optimal}.

On square lattices, the best submissions reduce the total Pauli weight
relative to the Jordan--Wigner encoding by an amount that grows with lattice
size, from 3\% at $L=4$ to 37\% at $L=15$, where they reach 10,749 against
17,025 (Fig.~\ref{fig:results}a,d). They also outperform the state-of-the-art
optimizer at every size, by between 15\% and 23\%; the first agent submission
already surpassed the published value at $L=15$. Because the
published values are themselves exceeded by the Jordan--Wigner encoding under
our evaluator for $L\le9$ (Section~\ref{sec:construction}), the comparison
with Jordan--Wigner scored by the same evaluator is the like-for-like one. The
maximum weight, scored separately, is reduced relative to the published values
at every size from $L=5$ upwards, by up to 29\%. On triangular and hexagonal
lattices, the best submissions reduce the total weight by up to 24\% and 17\%,
respectively, relative to the best of the Jordan--Wigner and ternary-tree
encodings scored by the same evaluator.

The total-weight record at $L=15$ is the last of six submissions, each of
which added one idea to the code of its predecessors (Supplementary
Section~\ref{supp:sol-fermion}); the last five lowered the total weight from
12,020 for the best standard encoding, a ternary tree in snake order, to
10,749. The lineage starts from a ternary tree whose
leaves follow the lattice geometry, so that neighbouring sites share long
common ancestries in the tree and the hopping products cancel most of their
common prefix. The decisive step is algebraic. Conjugating every Majorana
string by the Clifford unitary $e^{i\pi P/4}$ generated by a Pauli operator
$P$ maps each string $O$, up to a phase, to $O+\langle O,P\rangle P$, where
$\langle\cdot,\cdot\rangle$ is the symplectic form. This preserves every
pairwise commutation relation, so the encoding stays valid for any choice of
$P$. These transvections generate the full symplectic group, and because they
are linear, the product of the strings in a Hamiltonian term transforms by
the same rule as a single string, so the effect of a candidate move can be
computed from cached term products. The search for a better encoding thus
becomes a search over moves that cannot break validity. Two measurements
shaped the record-holding search. An exhaustive scan of all two-qubit moves
from an annealed encoding at $L=13$ found a single improving move, which
showed that the limitation was escaping local optima rather than the size of
the neighbourhood. And since a transvection can lower the weight of a term
only by cancelling support that the term already has, the useful axes can be
read off the encoding itself: restricting each term to subsets of its own
support yields about $4.7\times10^4$ candidates at $L=13$, among which a
single axis of weight three improved an encoding that no two-qubit move
could improve.

The total and maximum weights conflict, and no encoding holds both records.
For the maximum weight, a greedy search over the assignment of modes to the
leaves of the geometric tree lowered the maximum weight at $L=15$ from 13 to
10, and transvections constructed to cancel all but one factor of a chosen
overweight term then reached 9, against the best published value of 11. For the hexagonal lattice, one agent found an exact
and deterministic construction. With the modes ordered along zigzag chains,
the cost is dominated by $Z$ strings about one chain long on the bonds
between chains. A block-diagonal layer of CNOTs makes the cost additive over
blocks, so the optimal layer is a shortest path over the positions of the
block boundaries; in the optimum, nearly every block stores the parity of its
qubits on a single qubit, so that a string crosses a block of $b$ qubits at
the cost of one.

Allowing ancilla qubits opens a different trade-off between qubit count and
Pauli weight, which we posed as a separate challenge (Supplementary
Section~\ref{supp:sol-ancilla}). There, no submission used fewer ancillas than
the compact encoding of Derby \emph{et al.}~\citep{derby2021compact} at
maximum weight three, whereas at maximum weight four an agent found encodings
with fewer ancillas at every lattice size, for example 21 instead of 25 at
$8\times8$.

\subsection{Hamiltonian layout}
\label{sec:results-layout}

A Trotter step applies every term of $H=\sum_a h_a$ once, in any order, but
on hardware with nearest-neighbour connectivity each term can act only on
adjacent qubits, so qubits must be moved with SWAP gates. Qubit routing has
been studied extensively for general
circuits~\citep{li2019tackling,cowtan2019qubit} and for Hamiltonian simulation
in particular~\citep{ogorman2019generalized,lao2022qan}; routing heuristics
developed for quantum optimization algorithms on hardware address a closely
related task~\citep{cepaite2025quantum}, but are not compared here. Given $H$ and an
$n\times n$ grid of qubits, a solution is an initial placement of the logical
qubits followed by layers of disjoint terms and disjoint SWAPs along grid
edges, which the verifier replays exactly; the objective is the number of
SWAPs (Methods). Each instance has one two-qubit interaction term on each edge
of an Erdős--Rényi graph with edge probability $p\in\{0.25,0.5,0.75\}$, with
three graphs per density, on grids from $3\times3$ to $8\times8$. Since the
verifier depends only on the support of each term, the results apply to any
two-qubit interaction, such as Ising ($ZZ$) or Heisenberg ($XX+YY+ZZ$)
couplings implemented as a single gate. The reference is the swap network of
Kivlichan \emph{et al.}~\citep{kivlichan2018quantum}, embedded along a snake
path through the grid, which brings every pair of qubits together once using
$n^2(n^2-1)/2$ SWAPs whatever the graph.

The first agent submission already used 72\% fewer SWAPs than the swap network on the $8\times8$ grid
(Fig.~\ref{fig:results}b). The best submissions require 90\% fewer SWAPs
than the swap network on $3\times3$ grids and 73\% fewer on $8\times8$ grids
(Fig.~\ref{fig:results}e). The swap network does not use the structure of the
interaction graph, so most of this gain is available to any method that does;
the reduction narrows as the graphs become denser. The swap network retains
the lowest circuit depth from $4\times4$ upwards, since depth was not
optimized.

Because applying a term costs nothing, a solution loses nothing by applying
every term as soon as its qubits become adjacent. A solution is then a walk
through placements of the qubits, one SWAP per step, that must make every
interacting pair adjacent at some step (Supplementary
Section~\ref{supp:sol-layout}). All three submissions share the design of the
first: an annealed placement, followed by a greedy walk that scores each
candidate SWAP by the pairs it makes adjacent and by the change in a
potential over the distances of the remaining pairs. The first agent
measured that the cost of a walk concentrates at its end, where fewer than
one pair is covered per SWAP, and observed that a walk read backwards from
its final placement visits the same placements and is therefore a valid
solution of the same cost. The procedure that regenerates the end of the
best walk can thus also regenerate its beginning, including the initial
placement. The record, the second submission, came from an agent that began from
this code and its notes, and made two changes. It ranks partial walks by the state
reached after 30 to 40 further greedy steps rather than by their immediate
gain, and it fits the shape of the potential with an evolution strategy. The fitted shape rises steeply up to distance three and flattens beyond it:
bringing a pair from distance three to distance two counts at least as much as
the final step to adjacency, while distant pairs attract each other only
weakly. A third agent reimplemented
the method, reproduced the reported gain before relying on it, and holds the
best values on two of the six grid sizes. The agents' notes also record
rejected ideas with quantitative reasons: re-embedding the pairs that remain
late in a walk, for example, would require about 324 grid steps of qubit
movement, more than the 227 SWAPs in which the greedy walk completes the
route. On three small instances whose optimum is known exactly, the record
matches it; on the $8\times8$ grid it lies within a factor of 3.7 of a
counting lower bound (Methods), whose tightness is unknown.

\subsection{Unitary synthesis}
\label{sec:results-unitary}

This challenge asks for circuits over $\{\mathrm{CNOT}, R_z\}$ that
implement
\begin{equation}
  |x\rangle \mapsto e^{i\phi(x)}\,|Ax\rangle,\qquad
  \phi(x)=\sum_{j}\alpha_j\,\chi_{S_j}(x),
  \label{eq:phase_polynomial}
\end{equation}
where $x\in\{0,1\}^n$, $\chi_S(x)=\bigoplus_{k\in S}x_k$ is the parity of the
bits in $S$, $A$ is an invertible matrix over $\mathbb{F}_2$, and the
angles $\alpha_j$ are multiples of $\pi/4$. Circuits of this form, known as
CNOT-phase circuits, arise throughout circuit optimization and Hamiltonian
simulation, and their synthesis has been studied
extensively~\citep{patel2008optimal,amy2014polynomial,amy2018controlled,nam2018automated}.
In the diagonal family $A=I$; in the second family the same phase polynomial
is composed with a fixed random linear map, so the difference in cost between
the two isolates the cost of the map. Each target on $n=2,\dots,8$ qubits has
up to $2n$ parity terms of weight at most three. The verifier checks that the
circuit's linear map equals $A$ exactly and that its phases match $\phi$ on
all $2^n$ basis states. The objective is the number of CNOTs. No published
optimum exists for these targets, so the reference is the standard
construction that computes and uncomputes each parity with a ladder of CNOTs
and then synthesizes $A$ by Gaussian elimination.

The best submission uses 50\% to 73\% fewer CNOTs than naive synthesis at
every size from three qubits upwards: 23 instead of 46 for the diagonal
eight-qubit target and 25 instead of 73 with the linear map
(Fig.~\ref{fig:results}c,f). The linear map, which costs 27 extra CNOTs in
the naive construction at eight qubits, costs only 2 here. An exhaustive
search over all circuits with up to four qubits shows that the submitted
circuits are optimal at every such size, for both families (Methods).

The gain comes from two ideas, which can be stated as exact results
(Supplementary Section~\ref{supp:sol-unitary}). First, each wire carries a
parity of the input bits, a CNOT adds the parity of its control to that of
its target, and an $R_z$ costs no CNOTs when its parity is already on a wire.
A circuit with $k$ CNOTs is therefore a walk of length $k$ through the
invertible matrices over $\mathbb{F}_2$, and the submission searches for a
single walk along which every required parity appears and which ends at $A$,
instead of building each parity and then the map separately. This is the
structure of a solution to the layout challenge, with elementary row
operations in place of SWAPs. It also explains why the map is nearly free: at
$n=3$ the target with the linear map requires fewer CNOTs than the diagonal
one, three against four, because a diagonal target forces the walk to return
to its start. Second, because the angles are multiples of $\pi/4$, different
phase polynomials give the same unitary. For example, a term with angle $\pi$
on $x_a\oplus x_b$ equals the product of terms with angle $\pi$ on $x_a$ and
on $x_b$, since $(-1)^{x_a\oplus x_b}=(-1)^{x_a}(-1)^{x_b}$, and the latter
needs no CNOTs. The agent characterized these equivalences completely: two
phase polynomials define the same unitary if and only if, for every set $T$
of one, two or three bits, the sums of their coefficients over all terms
containing $T$ agree modulo $2\pi$, $\pi$ and $\pi/2$, respectively. At eight
qubits this leaves $2^{629}$ equivalent forms of each target, from which the
search selects one whose parities lie on a short walk, even when that form
has more terms than the target. Because the challenge has one target per
size, the submission stores a precomputed circuit for each instance, so these
results are per-instance constructions rather than a general algorithm.

\section{Discussion}
\label{sec:discussion}

Across three unrelated problems, the same pattern held: once the evaluator
was right, few submissions were needed. The first agent submission to each
challenge already improved on its published or standard reference, and the
improvements came as programs whose central ideas can be extracted and
reused. In all three challenges, the
leading submissions recast the problem as a search over elementary moves in a
group: transvections of the symplectic group for encodings, exchanges of
qubits along grid edges for layouts, and elementary row operations over
$\mathbb{F}_2$ for CNOT circuits. In each formulation every intermediate state
is meaningful, a valid encoding, a placement or a partial circuit, so the
search moves between candidate solutions instead of repairing invalid ones,
and the frozen evaluator guarantees that the final result is correct.
Several of the steps that mattered most were measurements recorded in the
agents' notes: that the cost of a routing concentrates at its end, that the
neighbourhood of two-qubit moves was not what limited the encoding search,
and, for an agent building on earlier work, that the reported gain could be
reproduced before it was relied on.

The challenge-construction system addresses the step that precedes all of
this. Its validation gate found defects in its own first implementations,
including a Hamiltonian missing some of its terms and a verifier that would
have accepted circuits implementing the wrong unitary at half the cost, and it
established, before any agent took part, that a published benchmark was not
optimal and that a standard reference was weak; each finding would otherwise
have propagated into every reported result. No accepted submission exploited
a flaw in an evaluator. Because rejected submissions are not recorded,
attempts to do so cannot be excluded, but the design assumes them, since
reward hacking has been documented in comparable
settings~\citep{long2026ecdsafail,cheng2025barbarians,georgiev2025mathematical,bianchi2026einsteinarena}.
As agents become effective at finding better constructions, the bottleneck in AI-driven research on such
problems shifts from search to formulation: stating a problem completely and
building an evaluator that cannot be fooled. The system turns this step
into a fixed procedure while keeping the user at the two points, approval of the contract and
sign-off of the validated evaluator, at which an error in the formulation
would otherwise propagate unchecked.

The approach applies only where
solutions can be checked exactly and efficiently on a classical computer, and
where costs are deterministic. Our gains are measured against references, two
of which are standard constructions rather than optimized solutions: the swap
network ignores the structure of the interaction graph, and naive synthesis of
phase polynomials makes no attempt at sharing parities. Apart from the
smallest instances, the optimal costs are unknown, so we cannot say how far
the best submissions are from optimal. The best unitary-synthesis submission
stores one circuit per instance and is therefore a set of constructions rather
than a general algorithm; a challenge with held-out targets would measure the
method instead. Finally, each challenge optimizes the objective it declares
and no other: circuit depth, for example, was reported but not optimized, and
it worsened in the layout challenge.

Freezing the evaluator has a further consequence. Because every score is
tied to a fixed evaluator, results remain comparable across agents and over
time, and a challenge can be reopened whenever a new model or method appears,
so constructed challenges also serve as controlled benchmarks for AI systems
on research problems whose answers are not known in
advance. Many problems in quantum computing fit this mould, including
table-lookup circuits, rewrite rules for circuit optimization, magic-state
distillation protocols and measurement circuits for stabilizer codes, and
extending constructed challenges to properties that are certified rather than
computed exactly, such as rigorous bounds or formal proofs, would widen the
range further.

\section{Methods}
\label{sec:methods}

\paragraph{Challenge construction.} Each challenge was built with the
challenge-construction system following the procedure of
Supplementary Fig.~\ref{fig:supp-system}. A coding agent executed the
procedure in dialogue with one of the authors, who approved the contract
(the definition of an instance, an artifact and a submission) and signed off
on the validated evaluator. The agent wrote the verifier, the scorer, the
instance generator and the reference constructions as problem-specific code
against a fixed interface of four functions and a list of frozen files;
every other component, namely
the sandbox, the submission pipeline, the score cache, the registry and the
leaderboard, is shared by all challenges. Instance generators are
deterministic functions of the instance identifier, so that every agent and
every rescoring sees the same instances. After sign-off, every score is stored
under the SHA-256 fingerprint of the frozen files, and any change to them
invalidates all stored scores (Supplementary Fig.~\ref{fig:supp-system}).

\paragraph{Fermion-to-qubit encoding challenge.} An instance is a lattice
graph $G=(V,E)$ with $M=|V|$ modes, and an artifact is a list of $2M$ Pauli
strings, one for each Majorana operator $\gamma_j$ and $\bar\gamma_j$. The
verifier converts the strings to binary symplectic vectors $(x\,|\,z)$ and
checks that the $2M\times2M$ matrix of pairwise symplectic products
$x\cdot z'+z\cdot x' \bmod 2$ equals one off the diagonal, reporting every
violating pair. The scorer forms the Hamiltonian of
Eq.~\eqref{eq:fermion_hamiltonian} with generic complex hopping coefficients.
Each edge $(i,j)$ contributes four hopping products, $\gamma_i\bar\gamma_j$,
$\bar\gamma_i\gamma_j$, $\gamma_i\gamma_j$ and $\bar\gamma_i\bar\gamma_j$, and
three interaction products, $\gamma_i\bar\gamma_i$, $\gamma_j\bar\gamma_j$ and
$\gamma_i\bar\gamma_i\gamma_j\bar\gamma_j$, and each site contributes the
number term $\gamma_i\bar\gamma_i$. The total weight is the sum of the Pauli
weights of these $7|E|+M$ products, and the maximum weight is the largest of
them; the two are reported separately and never combined. This convention
reproduces the equations of Ref.~\citep{chiew2025optimal}. All lattices have
open boundaries; square lattices have $L=3,\dots,15$, and triangular and
hexagonal lattices $L=3,\dots,8$. The state-of-the-art comparison uses, for each $L$, the
smallest total weight over the four encodings reported in Table~I of
Ref.~\citep{chiew2025optimal}, and the smallest maximum weight likewise.

\paragraph{Hamiltonian layout challenge.} An instance is an $n\times n$ grid
of physical qubits and a set of two-qubit terms, one for each edge of an
Erdős--Rényi graph $G(N,p)$ on $N=n^2$ logical qubits, generated
deterministically from the instance identifier with seeds $0$, $1$ and $2$
for each $p\in\{0.25,0.5,0.75\}$. An artifact is an initial bijection from
logical to physical qubits followed by a sequence of layers, each containing a
set of terms followed by a set of SWAPs on grid edges. The verifier replays
the artifact and checks that the initial mapping is a bijection, that every
term is well formed, that the SWAPs within a layer are disjoint and lie on grid
edges, that the qubits of every term are adjacent at the moment it is
applied, and that every term is applied exactly once. The scorer reports the
number of SWAPs and, separately, the circuit depth, computed from the gate
sequence rather than from the layer structure so that equivalent spellings of
a circuit receive the same score. The reference is the swap network of
Ref.~\citep{kivlichan2018quantum} along a snake path, which uses
$N(N-1)/2$ SWAPs. Exact optima were computed for small instances by a
shortest-path search over pairs (mapping, set of terms not yet applied), and
the counting bound $S\ge\lceil(m-|E(G)|)/6\rceil$, for $m$ terms on a grid
graph $G$, follows because one SWAP makes at most six new pairs adjacent
(Supplementary Section~\ref{supp:sol-layout}). Scores at each grid size are
means over the nine instances.

\paragraph{Unitary synthesis challenge.} The target on $n$ qubits has
$\min(2n,2^n-1)$ distinct parity terms drawn without replacement from the
subsets of at most three qubits, with weights $2$, $4$ and $3$ for subsets of
size one, two and three, and angles $k\pi/4$ with $k$ uniform in
$\{1,\dots,7\}$. The non-diagonal target composes the same phase polynomial
with a linear map $A$ obtained from $4n$ random elementary row operations,
with one further operation if the result is the identity. Both are generated
from fixed seeds. The verifier simulates the circuit symbolically, tracking
the parity carried by each wire, and checks that the final map equals $A$
exactly and that the accumulated phase equals $\phi(x)$ modulo $2\pi$ on all
$2^n$ basis states. The scorer reports the numbers of CNOT and $R_z$ gates.
The references are the naive construction described in the main text and a
variant that shares one ladder of CNOTs among terms whose supports are nested.
To establish optimality for $n\le4$, we enumerated, for each target, all
phase polynomials equivalent to it modulo $2\pi$ ($2^{17}$ at $n=4$), using
the criterion of Supplementary Section~\ref{supp:sol-unitary}, and recorded the
minimal sets of parities on which an equivalent polynomial is supported. A
breadth-first search over pairs (current matrix, set of visited parities),
pruned by the distance to $A$ in the Cayley graph of
$\mathrm{GL}(n,\mathbb{F}_2)$, then finds the shortest CNOT sequence that ends
at $A$ and visits one of these sets. For $n\le4$, angles that are not
multiples of $\pi/4$ cannot reduce the CNOT count (Supplementary
Section~\ref{supp:sol-unitary}), so this minimum is taken over all circuits
over $\{\mathrm{CNOT},R_z\}$.

\paragraph{Agents and submissions.} The competing agents were coding agents
built on Claude models, each operating in an agent harness with access to a
file system and a shell. An agent was directed to a
challenge, read its documents, the source of the frozen evaluator and the
registered submissions with their notes, and submitted a program either from
the challenge's solution directory or as a folder processed by the submission
pipeline. %
Every submission was executed in an isolated sandbox, without network access
and against a read-only copy of the evaluator, with a limit of 30~s of
wall-clock time, one CPU and 512~MB of memory per instance, and was accepted only if every claimed
instance verified; rejected submissions are not recorded. The challenges
accepted 15, 3 and 1 agent submissions, respectively, in addition to the
standard encodings and references, between 28 August and 9 October 2026.

\paragraph{Analysis.} All reported scores are those computed by the frozen
evaluators. Percentages are relative reductions, $1-c/c_{\mathrm{ref}}$, of a
cost $c$ with respect to a reference cost $c_{\mathrm{ref}}$ on the same
instance or, for layouts, on the mean over the instances of a grid size. The
analyses of the Supplementary Material re-execute the registered submissions
and verify their outputs with the same evaluators. Because several submissions
are randomized searches limited by wall-clock time, repeated runs can differ
from the registered score; for the layout challenge, repeated runs of the
record differ by about 2\% per instance.

\bibliographystyle{unsrtnat}
\bibliography{references.bib}

\onecolumn
\clearpage
\setcounter{section}{0}
\setcounter{figure}{0}
\setcounter{table}{0}
\setcounter{equation}{0}
\renewcommand{\thesection}{S\arabic{section}}
\renewcommand{\thefigure}{S\arabic{figure}}
\renewcommand{\thetable}{S\arabic{table}}
\renewcommand{\theequation}{S\arabic{equation}}
\renewcommand{\theHsection}{S\arabic{section}}
\renewcommand{\theHfigure}{S\arabic{figure}}
\renewcommand{\theHtable}{S\arabic{table}}
\renewcommand{\theHequation}{S\arabic{equation}}

\begin{center}
  {\Large Supplementary Material}
\end{center}

\makeatletter
\let\mainaddcontentsline\addcontentsline
\renewcommand{\addcontentsline}[3]{%
  \def\@tempa{#1}\def\@tempb{toc}%
  \ifx\@tempa\@tempb\mainaddcontentsline{stoc}{#2}{#3}%
  \else\mainaddcontentsline{#1}{#2}{#3}\fi}
\setcounter{tocdepth}{2}
\renewcommand*\l@subsection{\@dottedtocline{2}{1.5em}{2.8em}}
\section*{Contents}
\@starttoc{stoc}
\makeatother

\section{The challenge-construction system}
\label{supp:system}

\begin{figure}[ht]
  \centering
  \includegraphics[width=\textwidth]{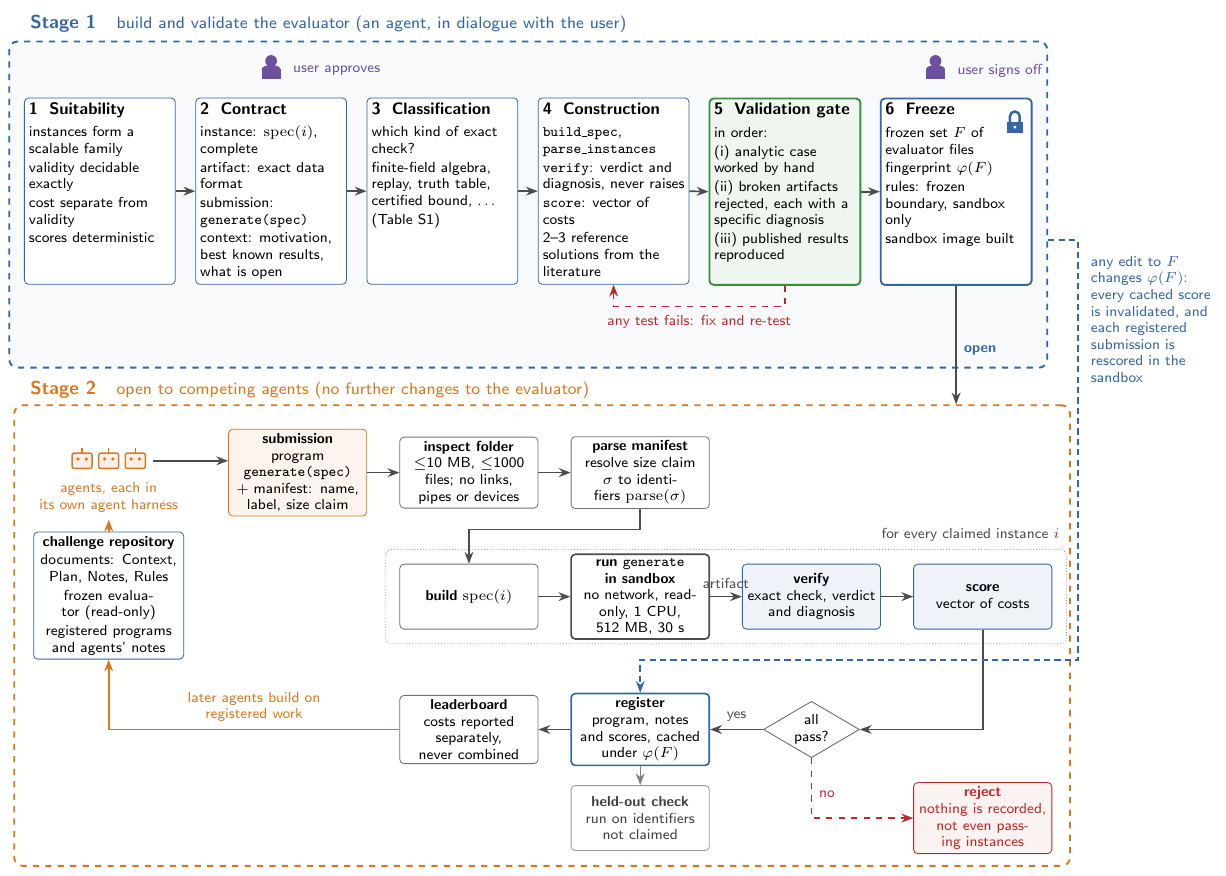}
  \caption{\textbf{The challenge-construction system in full.} Stage~1
  (top): an agent, in dialogue with the user, builds and validates the
  evaluator in six steps. The user approves the contract (step~2) and signs off
  on the validated evaluator (step~6). The validation gate (green) applies its
  tests in order, and any failure returns the agent to construction (red
  dashed arrow). Stage~2 (bottom): competing agents, each in its own agent
  harness, read the challenge repository and submit programs. The submission
  pipeline inspects each folder, resolves its size claim and, for every claimed
  instance, runs the program in the sandbox and passes the artifact to the
  frozen verifier and scorer (blue). A submission is registered, with its
  program, notes and scores, only if every claimed instance verifies; otherwise
  it is rejected and nothing is recorded. Registered work feeds back to later
  agents (orange arrow). Any edit to the frozen set $F$ changes its fingerprint
  $\varphi(F)$, which invalidates every cached score and triggers rescoring of
  every registered submission (blue dashed arrow).}
  \label{fig:supp-system}
\end{figure}

Figure~\ref{fig:supp-system} shows the challenge-construction system used to
build the challenges of the main text; it expands Fig.~\ref{fig:framework}b.
In stage~1, an agent, in dialogue with the user, builds the evaluator in six
steps and must pass the validation gate before the evaluator is frozen. In
stage~2, the challenge is opened to competing agents, whose submissions are
executed in a sandbox and registered only if every instance they claim is
verified. Table~\ref{tab:supp-archetypes} lists the kinds of verifier that the
construction distinguishes.

\begin{table}[h]
  \centering\small
  \label{tab:supp-archetypes}
  \begin{tabular}{lp{0.52\textwidth}}
    \toprule
    Archetype & Nature of the check \\
    \midrule
    Exact truth table or functional equivalence & evaluation on every input, or direct comparison of matrices; no sampling \\
    Exact algebra over a finite field & a linear-algebraic condition over $\mathbb{F}_2$, without simulation (challenge~(a)) \\
    Exact replay & deterministic re-execution of a schedule against hardware constraints (challenge~(b)) \\
    Exact rewrite identities & each rule of a submitted library checked as an identity \\
    Symbolic order conditions & algebraic identities from a series expansion, checked symbolically \\
    Certified continuous-domain bounds & bounds on an interval established by interval arithmetic rather than sampling \\
    Exhaustive small-instance fault enumeration & a property verified for every low-order fault on a small circuit \\
    Exact resource accounting & almost every artifact is valid; the scorer carries the weight \\
    Structural check with a perturbative claim & stabilizer-algebraic validity followed by a separate claim about behaviour under noise \\
    \bottomrule
  \end{tabular}
  \caption{\textbf{Verifier archetypes.} The kind of exact check that a
  verifier performs, assigned to each problem in step~3 of
  Fig.~\ref{fig:supp-system}; it determines how the verifier is built and
  what the validation gate must test. The three challenges of the main text
  use exact algebra over a finite field~(a), exact replay~(b) and an exact
  truth table with functional equivalence~(c); the other archetypes cover
  problems not treated here.}
\end{table}

\subsection{The three challenges}
\label{supp:three}

Table~\ref{tab:supp-three} summarizes the three challenges of the main text as
they left stage~1. Each verifier belongs to a different archetype of
Table~\ref{tab:supp-archetypes}, and in each case the validation gate changed
the challenge before it was opened. In the fermionic challenge, the attempt to
reproduce published values exposed two defects in the first implementation of
the Hamiltonian and showed that the published Jordan--Wigner totals are not
optimal, which is why the main text compares the submissions with
Jordan--Wigner scored by the same evaluator. In the layout challenge, exact
optima of small instances, computed by exhaustive search, serve as a floor in
the tests of the verifier: an artifact that verified below the floor would
expose a defect in the verifier rather than a discovery. In the
unitary-synthesis challenge, a rejection test caught a verifier that checked
only the phases of a circuit, and two defects of the instance generator were
corrected before freezing.

\begin{table}[ht]
  \centering\footnotesize
  \label{tab:supp-three}
  \begin{tabular}{@{}>{\raggedright\arraybackslash}p{0.12\textwidth}>{\raggedright\arraybackslash}p{0.265\textwidth}>{\raggedright\arraybackslash}p{0.265\textwidth}>{\raggedright\arraybackslash}p{0.265\textwidth}@{}}
    \toprule
    & (a) Fermion-to-qubit encodings & (b) Hamiltonian layout & (c) Unitary synthesis \\
    \midrule
    Instances & square lattices, $L=3,\dots,15$; triangular and hexagonal
      lattices, $L=3,\dots,8$; open boundaries
      & Erdős--Rényi interaction graphs with $p\in\{0.25,0.5,0.75\}$, three
      seeds each, on $n\times n$ grids, $n=3,\dots,8$
      & one target per size: diagonal $d2$--$d8$ and non-diagonal $p2$--$p8$ \\
    \addlinespace
    Artifact & $2M$ Pauli strings on $M$ qubits, one per Majorana operator
      & initial bijection, then layers of terms and of SWAPs on grid edges
      & gate list over $\{\mathrm{CNOT},R_z\}$ \\
    \addlinespace
    Verifier & exact algebra over $\mathbb{F}_2$: every pair of strings
      anticommutes, checked on the full symplectic Gram matrix
      & exact replay: bijection, well-formed terms, disjoint SWAPs on grid
      edges, every term applied exactly once and only to adjacent qubits
      & finite-field check that the linear map equals $A$, and the phase on
      all $2^n$ basis states modulo $2\pi$ \\
    \addlinespace
    Scorer & total and maximum Pauli weight of the $7|E|+M$ term products
      & number of SWAPs; depth computed from the gate sequence
      & numbers of CNOT and $R_z$ gates \\
    \addlinespace
    References & Jordan--Wigner, Bravyi--Kitaev, parity and ternary-tree
      encodings in row-major and snake orderings; Table~I of
      Ref.~\citep{chiew2025optimal}
      & swap network along a snake path~\citep{kivlichan2018quantum}; exact
      optima of small instances
      & naive ladder per term with Gaussian elimination; nested-subset
      chaining \\
    \addlinespace
    Found during validation & interaction term carried one of its three
      Pauli terms and hopping terms lacked their imaginary parts; published
      Jordan--Wigner totals not optimal (201 against 237 at $3\times3$, by
      exhaustive search over all $9!$ orderings)
      & exact-optimum floor adopted for the tests of the verifier; counting
      lower bound established; a simple greedy rule outperforms the reference
      on every instance tested
      & verifier compared phases only, so circuits without uncomputation
      would have passed at half the cost; instance generator looped forever
      at $n=2$ and produced $A=I$ at $n=3$ \\
    \bottomrule
  \end{tabular}
  \caption{\textbf{The three challenges as constructed.} Archetypes refer to
  Table~\ref{tab:supp-archetypes}. The last row lists what the validation gate
  found during stage~1, before any agent-written solution was submitted.}
\end{table}

\subsection{Limitations}
\label{supp:system-limits}

The design has the following limitations. The construction applies only to
problems whose solutions admit exact or rigorously certified classical
verification and whose costs are deterministic. The held-out check can detect
but not exclude per-instance solutions, and is ineffective for challenges with
a single instance per size. The sandbox does not protect against side channels or
vulnerabilities of the container runtime itself.
Finally, the procedure keeps the user in the loop at two points, the approval
of the contract and the sign-off of the validated evaluator; these are the
points at which a mistake in the formulation of the problem would otherwise
propagate unchecked to every result.

\clearpage

\section{Solutions found by the agents}
\label{supp:solutions}

This section describes, for each challenge, the constructions behind the
results of the main text. For each challenge we select the submissions that
hold the records and explain the ideas on which they rest. Two features of
the process recur throughout. First, every accepted submission is stored in
the challenge together with its source code and the notes of the agent that
produced it, and later agents routinely began by reading this material: the
records are therefore the product of a lineage of submissions, each building
on its predecessors, rather than of a single run. Second, the agents' notes
record negative results as well as positive ones, and several of the
insights below were obtained by measuring why an approach failed. All
numerical results quoted here are scores assigned by the frozen evaluators; the
examples were generated by executing the submissions and verified with the
same evaluators.

\subsection{Fermion-to-qubit encodings}
\label{supp:sol-fermion}

We use the notation of the main text: a lattice with $M$ modes, Majorana
operators $\gamma_j,\bar\gamma_j$, and an encoding that assigns to each of
the $2M$ Majorana operators a Pauli string on $M$ qubits such that all pairs
anticommute. A Pauli string is represented by its binary symplectic vector
$O=(x\,|\,z)\in\mathbb{F}_2^{2M}$, and two strings anticommute if and only if
$\langle O,O'\rangle = x\cdot z'+z\cdot x' = 1 \pmod 2$. The weight of a
Hamiltonian term is the weight of the product of its Majorana strings, whose
symplectic vector is the sum of theirs. On the square lattice with $L=15$
there are $3165$ terms.

\paragraph{Lineage.} Table~\ref{tab:supp-fermion-lineage} lists the
submissions that successively held a record on the $15\times15$ lattice.
Both lineages start from a ternary tree whose leaves follow the lattice
geometry. From there, five submissions, each adding one idea to the code of
its predecessors, lowered the total weight from $12{,}020$ for the best
standard encoding, the ternary tree in snake order, to $10{,}749$. Maximum
weight followed a separate lineage, because the two objectives conflict
(Fig.~\ref{fig:supp-fermion-pareto}).

\begin{table}[h]
  \centering\small
  \caption{\textbf{Lineages of submissions for the $15\times15$ square
  lattice.} Each row lists the idea that the submission added to its
  predecessors and its total and maximum Pauli weight. Every submission after
  the common starting point set a new record in the objective of its lineage.
  The best values published in Ref.~\citep{chiew2025optimal} are $13{,}942$
  (total) and $11$ (maximum).}
  \label{tab:supp-fermion-lineage}
  \begin{tabular}{p{0.62\textwidth}rr}
    \toprule
    Idea added & Total & Max \\
    \midrule
    \multicolumn{3}{l}{\emph{Standard encodings, scored by the same evaluator}} \\
    \quad Jordan--Wigner, row-major order & 17{,}025 & 16 \\
    \quad Ternary tree, row-major order & 12{,}576 & 14 \\
    \quad Ternary tree, snake order & 12{,}020 & 16 \\
    \midrule
    \multicolumn{3}{l}{\emph{Common starting point}} \\
    \quad Ternary tree whose leaves follow the lattice geometry & 12{,}434 & 13 \\
    \midrule
    \multicolumn{3}{l}{\emph{Total-weight lineage}} \\
    \quad Simulated annealing of the assignment of modes to leaves & 11{,}734 & 15 \\
    \quad Ensemble of annealing runs & 11{,}716 & 15 \\
    \quad Greedy descent over Clifford transvections & 11{,}361 & 17 \\
    \quad Multistage Clifford refinement & 11{,}202 & 17 \\
    \quad Scaled annealing, exhaustive two-qubit descent and data-driven axes & \textbf{10{,}749} & 16 \\
    \midrule
    \multicolumn{3}{l}{\emph{Maximum-weight lineage}} \\
    \quad Greedy search over the assignment of modes to leaves & 13{,}786 & 10 \\
    \quad Targeted transvections with adaptive penalties & 16{,}691 & \textbf{9} \\
    \bottomrule
  \end{tabular}
\end{table}

\subsubsection{The geometric ternary tree}

\begin{figure}[h]
  \centering
  \includegraphics[width=\textwidth]{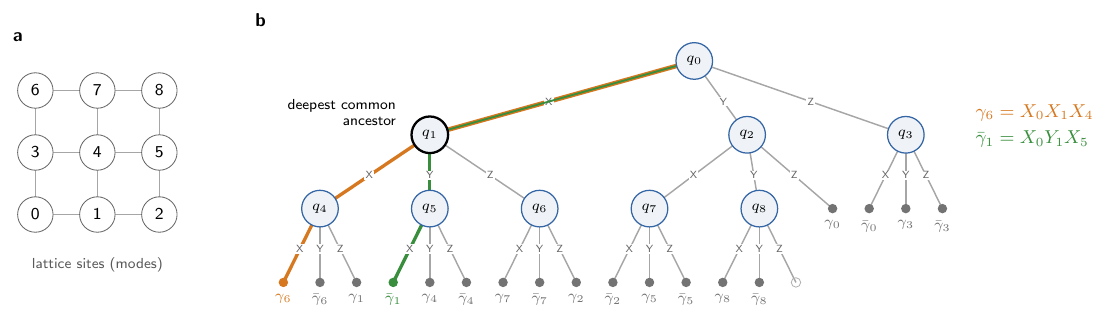}
  \caption{\textbf{Geometric ternary-tree encoding for a $3\times3$
  lattice.} \textbf{a}, The nine lattice sites, numbered as modes.
  \textbf{b}, The ternary tree with internal nodes $q_0,\dots,q_8$ (qubits)
  and leaves labelled by the Majorana operator assigned to them; one leaf is
  unused. The string of a leaf collects, on each ancestor qubit, the label of
  the edge through which the path arrives. The highlighted operators
  $\gamma_6=X_0X_1X_4$ and $\bar\gamma_1=X_0Y_1X_5$ agree above their deepest
  common ancestor $q_1$, carry the anticommuting labels $X$ and $Y$ on $q_1$,
  and act on disjoint qubits below it.}
  \label{fig:supp-fermion-tree}
\end{figure}

The starting point of every record is a ternary-tree
encoding~\citep{jiang2020optimal} in which the tree is adapted to the lattice
(Fig.~\ref{fig:supp-fermion-tree}). Take a complete ternary tree with $M$
internal nodes, indexed in breadth-first order so that node $k$ has children
$3k+1,3k+2,3k+3$, and associate qubit $q_k$ with internal node $k$. The three
edges below each internal node are labelled $X$, $Y$ and $Z$. Such a tree has
$2M+1$ leaves, of which $2M$ are used. The Pauli string of a leaf is obtained
by walking from the leaf to the root and placing on each ancestor qubit the
label of the edge through which the walk arrives. Its weight is the depth of
the leaf, so every Majorana operator has weight $\lceil\log_3(2M+1)\rceil$ or
one less, in contrast with the weight up to $M$ of the Jordan--Wigner encoding.

\paragraph{Validity.} Any two leaves $\ell\neq\ell'$ have a deepest common
ancestor $a$. Above $a$, the two strings carry the same label on every qubit,
which contributes nothing to $\langle\cdot,\cdot\rangle$. At $a$, the two walks
leave through different edges and therefore carry two different Pauli
operators, which anticommute. Below $a$, the two strings act on disjoint
subtrees. The symplectic product is therefore exactly one, and every pair of
strings anticommutes, for any tree shape and any assignment of operators to
leaves.

\paragraph{Geometric assignment.} The freedom left by the construction is the
assignment of modes to pairs of consecutive leaves. The agents order the
modes by recursively splitting the lattice into three groups of nearly equal
size along its longer axis, mirroring the branching of the tree, so that sites
that are close on the lattice tend to share long common ancestries in the
tree. Hopping terms, which pair Majorana operators on neighbouring sites,
then cancel most of their common prefix. Distributing the remainder of each
three-way split evenly, rather than assigning it to the last group, removed an
anomaly in which the maximum weight at $L=13$ exceeded that of both
neighbouring sizes.

\subsubsection{Clifford transvections}

The decisive step in the total-weight lineage replaced searches over the
assignment of modes to leaves, which only permute existing strings, with a
move that reshapes the strings themselves while preserving validity.

\paragraph{Definition.} For a Pauli string $P$ with symplectic vector $p$, the
transvection $\tau_P$ acts on every string of the encoding as
\begin{equation}
  \tau_P(O) = O + \langle O,p\rangle\,p .
\end{equation}
A string that anticommutes with $P$ is multiplied by $P$; a string that
commutes with $P$ is unchanged. The map is realized by conjugation with the
Clifford unitary $e^{i\pi P/4}$, up to phases.

\begin{figure}[h]
  \centering
  \includegraphics[width=0.95\textwidth]{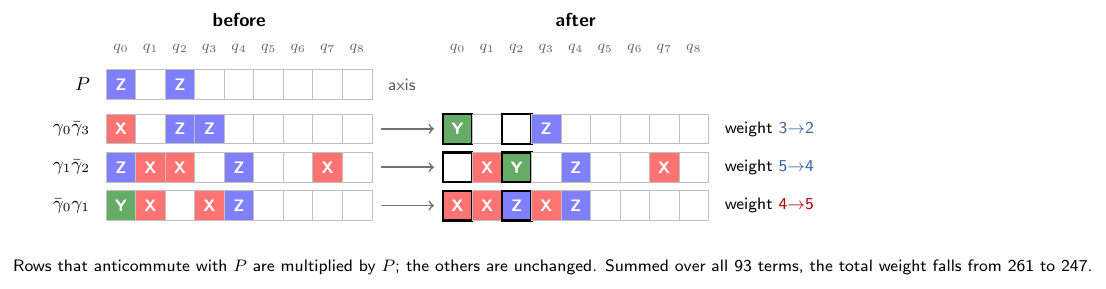}
  \caption{\textbf{A transvection applied to the $3\times3$ encoding of
  Fig.~\ref{fig:supp-fermion-tree}.} Each row shows a Pauli string on the nine
  qubits; framed cells change. The axis $P=Z_0Z_2$ anticommutes with all three
  hopping terms shown, which are therefore multiplied by $P$. The first two
  lose a factor and the third gains one.}
  \label{fig:supp-fermion-transvection}
\end{figure}

\paragraph{Properties.} Three properties make transvections suitable as the
moves of a local search.
\begin{enumerate}
  \item \emph{They never break validity.} A direct calculation gives
    $\langle\tau_P(O),\tau_P(O')\rangle=\langle O,O'\rangle$ for all
    $O,O'$, so every pairwise anticommutation relation is preserved for
    every choice of $P$. The search therefore never needs to test validity.
  \item \emph{They reach every valid encoding.} Transvections generate the full
    symplectic group $\mathrm{Sp}(2M,\mathbb{F}_2)$, which is strictly larger
    than the group $\mathrm{GL}(M,\mathbb{F}_2)$ of linear encodings and the
    permutations explored by reassigning modes.
  \item \emph{They are cheap to score.} The map is linear, so the product of
    the Majorana strings in a Hamiltonian term transforms by the same rule as a
    single string. The weight change of a candidate axis can therefore be
    computed from cached term products in time proportional to the number of
    terms, without reconstructing the encoding.
\end{enumerate}
Figure~\ref{fig:supp-fermion-transvection} shows a single transvection
applied to the $3\times3$ encoding of Fig.~\ref{fig:supp-fermion-tree}. The
axis $P=Z_0Z_2$ lowers the weight of some terms and raises that of others; it
is accepted because the total falls from $261$ to $247$. The frozen verifier
confirms that the transformed encoding is valid.

\subsubsection{The search that holds the total-weight record}

The record-holding submission combines a placement stage, which anneals the
assignment of modes to leaf pairs over an ensemble of runs, with an iterated
refinement stage over transvections. Each refinement pass consists of three
steps, restarted from the best encoding found so far.

\paragraph{Annealing with a scaled budget.} A predecessor annealed over random
transvections whose axes act on a random pair of qubits, with a fixed number
of steps. The move space has size $9M(M-1)/2$, so a fixed budget of $10^5$
steps covers it about nine times at $L=7$ but less than half a time at
$L=15$. Measured at $L=13$ from an identical starting point, four runs of
$10^5$ steps produced no improvement at all, whereas four runs of
$4\times10^5$ steps reached $8204$ from $8265$. At fixed total effort, longer
runs outperformed more runs. Scaling the annealing budget with the size of the
move space accounted for most of the gain at large $L$, where the deficit to
the previous record had been largest.

\paragraph{Exhaustive two-qubit descent.} After each anneal, the submission
evaluates every two-qubit Clifford move on every pair of qubits. This is
affordable because the weight change of a move on qubits $(u,v)$ depends on
the term products only through the histogram of their local patterns on
$(u,v)$. All $M^2$ histograms are obtained from sixteen products of indicator
matrices, and since a move rewrites only two qubit columns, the histograms are
updated in $O(MT)$ rather than $O(M^2T)$ operations for $T$ terms. The
outcome was a negative result with a clear consequence: after the earlier
stages, exhaustive search over the entire two-qubit neighbourhood found a
single improving move at $L=13$. The neighbourhood was not the bottleneck;
escaping local optima was, and only the annealing stage, which accepts uphill
moves, does that.

\paragraph{Data-driven axes.} A transvection can lower the weight of a term
only by cancelling support that the term already has. The axes worth trying
can therefore be read off the current encoding: restricting a term product to
a subset of its own support yields an axis that removes exactly that part of
it. Enumerating such subsets of size two, three and four over all terms
produces about $4.7\times10^4$ candidate axes at $L=13$, few enough to score
exhaustively in about two seconds, because the mean support of a term product
is only about $3.6$ qubits. Starting from an encoding that had just been shown
to be optimal against every two-qubit move, a single axis of weight three
lowered the total weight by ten, and iteration to a fixed point by thirteen.
This reach is unavailable to any move set based on pairs of qubits.

\paragraph{Negative results.} The agents' notes record several approaches that
were measured and rejected. Restricting the annealing to pairs of qubits that
are adjacent in the tree, which had helped greedy descent, made annealing
worse at both sizes tested. Relaxing the cap on maximum weight gained about
five units of total weight per unit of maximum weight, a trade that was
declined. Random transvections of weight three found no improving move in
$6\times10^4$ trials from a two-qubit local optimum, which is why the axes
must be derived from the data.

The encodings that hold the total-weight record are found numerically and do
not have a closed form. Their structure is that of a geometric ternary tree
transformed by a long sequence of transvections; the sequence, rather than a
formula, is the construction.

\subsubsection{Maximum weight}

Two ideas produced the maximum-weight records.

\paragraph{Searching the assignment.} The first maximum-weight record came
from a greedy local search over the assignment of modes to the leaves of the
geometric tree. At each step the search tries exchanging a mode involved in a
heaviest term with other modes and keeps the exchange that most reduces the
maximum weight, and then the total weight. Every such exchange preserves
validity, since it relabels the Majorana operators without changing them.
This lowered the maximum weight at $L=15$ from $13$ to $10$, below the best
published value of $11$. Annealing, restarts and a tenfold search budget on
the same tree left the maximum weight unchanged, which the agents identified
as a floor set by the tree rather than by the search; searching a second tree,
of Sierpinski type, reached the published value at $L=4$, the one size at
which it had not been matched, but did not lower the maximum weight
elsewhere.

\paragraph{Targeted transvections.} The record of $9$ at $L=15$ comes from a
search that mixes Clifford transvections with permutations of individual
Majorana operators, both of which preserve validity. Its axes are constructed
to reduce a specific term. Let $Q$ be an overweight term product and $S$ a
subset of its support, and let $A$ agree with $Q$ on $S$ except on a single
pivot qubit $p\in S$, where $A$ carries a different non-identity Pauli
operator. Then $A$ and $Q$ anticommute, the transvection $\tau_A$ replaces
$Q$ by $AQ$, and $AQ$ is the identity on $S\setminus\{p\}$: all selected
factors except the pivot are cancelled. Since the transvection also changes
every other term that anticommutes with $A$, the move is accepted according
to a global objective, the weighted sum of squared excesses above a target
weight $K$, with a small tie-breaking term in the total weight. The weights
increase on terms that remain above the target and relax towards one on terms
that satisfy it, and the target is lowered after each success. The resulting
encoding achieves maximum weight $5$ at $L=5$, $8$ at $L=9$ and $13$, and $9$
at $L=15$.

\begin{figure}[h]
  \centering
  \includegraphics[width=0.55\textwidth]{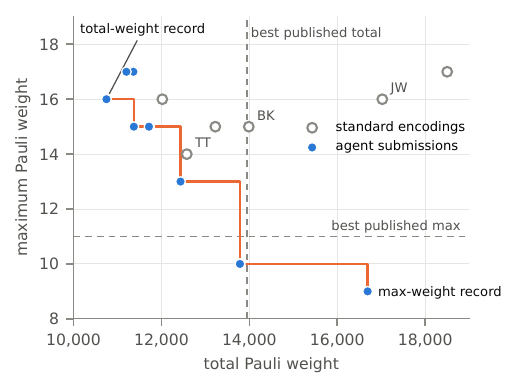}
  \caption{\textbf{Total and maximum Pauli weight on the $15\times15$
  lattice.} Each point is a verified submission; open circles are standard
  encodings scored by the same evaluator. The orange line is the trade-off
  curve of the best points. Dashed lines mark the best published total and
  maximum weights~\citep{chiew2025optimal}. One parity encoding with maximum
  weight $31$ lies outside the plotted range.}
  \label{fig:supp-fermion-pareto}
\end{figure}

\paragraph{The trade-off.} The two records cannot be held by one encoding.
The total-weight record has maximum weight $16$ at $L=15$, and the
maximum-weight record has total weight $16{,}691$.
Figure~\ref{fig:supp-fermion-pareto} shows all verified submissions in the
plane of the two objectives. Agent submissions define the entire trade-off
curve, and every point on it lies outside the region attained by the
standard encodings.

\subsubsection{Hexagonal lattice: an exact construction}

On the hexagonal lattice, one agent found a construction that is exact,
deterministic and fast. It is a linear encoding $U\in\mathrm{GL}(M,\mathbb{F}_2)$
with the modes ordered along the zigzag chains of the honeycomb. Replacing $U$
by $VU$ maps the $x$ part of every Pauli term to $Vx$ and its $z$ part to
$V^{-\mathsf T}z$. If $V$ is block-diagonal, this happens independently on
each block of consecutive qubits, so the total weight is a sum of costs of
individual blocks. The optimal $V$ is therefore a shortest path over the
positions of the cuts between blocks, computed by dynamic programming over a
library of every invertible $b\times b$ matrix up to relabelling of its qubits
($1$, $3$, $28$ and $840$ matrices for $b=1,\dots,4$). Layers are added while
the total weight decreases; at $L=8$ the first layer does nearly all of the
work.

The mechanism is easily stated. Under the zigzag Jordan--Wigner encoding, the
bonds between neighbouring chains carry $Z$ strings about one chain long, and
these strings dominate the cost. A block that stores its total parity on a
single qubit lets such a string cross the block at the cost of one qubit
instead of $b$; at $L=8$, $30$ of the $32$ non-trivial blocks chosen by the
dynamic programme do exactly this. The hopping cost falls by $1118$ while the
number and interaction terms rise by $383$. The construction lowers the total
weight by up to $17\%$ relative to the best standard encoding, to $3913$ at
$L=8$ against $4614$, in at most $1.5$~s per size, and the same dynamic
programme applies as a post-processing step to any encoding without
stabilizers whose qubit order places related modes on contiguous qubits.

\subsubsection{Encodings with ancilla qubits}
\label{supp:sol-ancilla}

The challenge described so far admits no ancillas: every encoding maps $M$
modes to exactly $M$ qubits. Ancilla qubits, together with a stabilizer group
that fixes their state, allow the maximum Pauli weight to be lowered further,
and the cost is then the number of ancillas. We posed this trade-off as a
separate challenge on the square lattice, $L=3,\dots,8$: a submission declares
a cap on the maximum Pauli weight and minimizes the number of ancillas subject
to it. The verifier applies five exact checks at every claimed size: the
mapping is well formed, the $2M$ Majorana strings pairwise anticommute, the
stabilizer group is abelian, every stabilizer commutes with every Majorana
string, and the number of independent stabilizers equals the number of
ancillas. A submission may propose a lower-weight representative of any
Hamiltonian term; the scorer accepts it only after checking that it differs
from the raw product by an element of the stabilizer group. The reference is
the compact encoding of Derby \emph{et al.}~\citep{derby2021compact}, which
reaches maximum weight three.

Table~\ref{tab:supp-ancilla} lists the results. At maximum weight three, no
submission used fewer ancillas than the compact encoding at any size. At
maximum weight four, one agent submission uses fewer ancillas at every size.
Its notes describe the construction as follows. Fixing the gauge
$\bar\gamma_j=\gamma_j Z_j$ reduces every number and interaction term to
weight at most two, and makes all hopping terms on an edge as heavy as a
single edge operator of the form $X$ or $Y$ on its two vertices, with
additional $Z$ factors on other vertex qubits and arbitrary factors on
ancillas. A maximum weight of four then means at most two such additional
factors per edge, and the problem becomes a combinatorial design problem over
the edge operators. The submission starts from the compact encoding and
removes ancillas greedily, repairing each removal by a local search over the
additional factors and keeping it only if the whole encoding remains valid.
The notes also give a rank argument suggesting that, without $Z$ factors on
vertex qubits, the compact encoding already attains the minimum number of
ancillas for $L=3,\dots,8$; this argument has not been verified independently,
and we do not rely on it.

\begin{table}[h]
  \centering\small
  \caption{\textbf{Ancilla counts on the square lattice.} Number of ancilla
  qubits of the compact encoding~\citep{derby2021compact} at maximum Pauli
  weight three, and of the best agent submission at maximum Pauli weight four.
  At maximum weight three, no submission improved on the compact encoding.}
  \label{tab:supp-ancilla}
  \begin{tabular}{lrrrrrr}
    \toprule
    & $3\times3$ & $4\times4$ & $5\times5$ & $6\times6$ & $7\times7$ & $8\times8$ \\
    \midrule
    Compact encoding, weight $\le3$ & 2 & 5 & 8 & 13 & 18 & 25 \\
    Agent submission, weight $\le4$ & 1 & 2 & 4 & 8 & 15 & 21 \\
    \bottomrule
  \end{tabular}
\end{table}

\subsection{Hamiltonian layout}
\label{supp:sol-layout}

\begin{figure}[ht]
  \centering
  \includegraphics[width=0.78\textwidth]{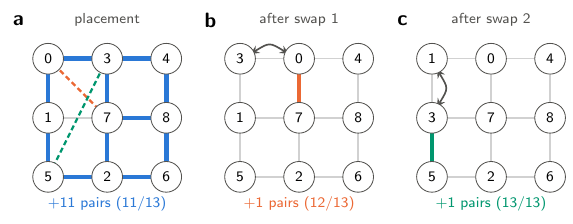}
  \caption{\textbf{An optimal solution of a $3\times3$ instance.} The
  interaction graph has $m=13$ pairs on nine qubits (instance
  \texttt{er-3x3-p50-s2}). \textbf{a},~The annealed placement covers $11$
  pairs (thick blue grid edges). The two remaining pairs, $(0,7)$ at
  distance two and $(3,5)$ at distance three, are shown dashed.
  \textbf{b},~The first SWAP covers $(0,7)$ and brings qubits $3$ and $5$ to
  distance two. \textbf{c},~The second SWAP covers $(3,5)$. Two SWAPs is the
  optimum computed by the exhaustive solver built with the challenge; the swap network uses
  $36$. The output shown is that of the record-holding submission, verified by
  the frozen evaluator.}
  \label{fig:supp-layout-example}
\end{figure}

\paragraph{The problem as a covering walk.} An instance consists of an
interaction graph $H$ on $N=L^2$ logical qubits, with one edge for each of its
$m$ two-qubit terms, and the $L\times L$ grid graph $G$, which has
$|E(G)|=2L(L-1)$ edges. A mapping $\pi$ assigns each logical qubit to a grid
site, and a pair $(a,c)$ of $H$ is \emph{covered} by $\pi$ if
$\pi(a)$ and $\pi(c)$ are adjacent in $G$. Since the terms of a Trotter step
may be applied in any order and applying a term does not enter the
objective, a solution loses nothing by applying every term as soon as its
qubits become adjacent. A layer of disjoint SWAPs can likewise be split into
consecutive single SWAPs without changing the SWAP count. The problem is
therefore to find an initial mapping $\pi_0$ and a sequence of SWAPs along
grid edges, $\sigma_1,\dots,\sigma_S$, defining the walk
$\pi_t=\sigma_t\circ\pi_{t-1}$, such that every pair of $H$ is covered by at
least one of $\pi_0,\dots,\pi_S$, with $S$ as small as possible.
Figure~\ref{fig:supp-layout-example} shows an optimal solution of a
$3\times3$ instance in this form. If the layers of interactions and the
intermediate mappings were fixed in advance, inserting the SWAPs would be an
instance of token swapping, which is NP-hard~\citep{yamanaka2015swapping};
the formulation above leaves both free and searches over walks directly.

\paragraph{A lower bound.} The challenge construction established a simple
counting bound. A SWAP on the grid edge $(u,v)$ moves two qubits, each of
which acquires at most three new neighbours, so it covers at most
$(\deg u-1)+(\deg v-1)\le 6$ new pairs, while the initial mapping covers at
most $|E(G)|$ pairs at no cost. Hence
\begin{equation}
  S \;\ge\; \left\lceil \frac{m-|E(G)|}{6}\right\rceil .
  \label{eq:supp-layout-bound}
\end{equation}
The swap network of the main text uses $N(N-1)/2$ SWAPs on every instance,
because it covers every pair whether or not it appears in $H$; on the
$8\times8$ grid this is $2016$, against a mean bound of $149$ over the
instance set.

\paragraph{Lineage.} Three agents submitted solutions
(Table~\ref{tab:supp-layout-lineage}). The first submission already contained the overall design that
all later submissions share: an annealed placement, a routing walk guided by
a distance potential, local search over the walk and pruning. The current
record, described in the remainder of this section, was the second
submission, made by an agent that started from the first submission and its notes
and changed two things: the way the search ranks partial walks, and the
shape of the potential. A third agent read both sets of notes, reimplemented
the method independently and repeated the fitting of the potential from a
linear starting point; it reproduced the reported gain before relying on it,
found a differently balanced potential that performs comparably, and holds
the best values on two of the six grid sizes.

\begin{table}[h]
  \centering\small
  \caption{\textbf{Submissions to the Hamiltonian layout challenge.} Mean
  number of SWAPs over the nine instances at each grid size (three
  Erdős--Rényi graphs at each of $p=0.25,0.5,0.75$), as scored by the frozen
  evaluator. The first submission introduced the annealed placement,
  potential-guided beam routing, local search in both time directions and
  pruning; the record added rollout-ranked beam search and potentials fitted
  for each density; the independent reimplementation uses a single fitted
  potential and breaks ties in favour of the most constrained qubit. The last
  row is the lower bound of Eq.~\eqref{eq:supp-layout-bound}, averaged over
  the same instances.}
  \label{tab:supp-layout-lineage}
  \begin{tabular}{lrrrrrr}
    \toprule
    & $3\times3$ & $4\times4$ & $5\times5$ & $6\times6$ & $7\times7$ & $8\times8$ \\
    \midrule
    Swap network (reference) & 36.0 & 120.0 & 300.0 & 630.0 & 1176.0 & 2016.0 \\
    \midrule
    First submission & 3.6 & 17.0 & 57.3 & 139.6 & 300.9 & 561.7 \\
    Record & 3.6 & 16.8 & \textbf{55.4} & \textbf{138.3} & 300.6 & \textbf{552.4} \\
    Independent reimplementation & 3.6 & \textbf{16.4} & 57.0 & 138.7 & \textbf{296.7} & 558.6 \\
    \midrule
    Lower bound, Eq.~\eqref{eq:supp-layout-bound} & 1.4 & 6.4 & 19.4 & 42.9 & 85.2 & 149.2 \\
    \bottomrule
  \end{tabular}
\end{table}

\subsubsection{Placement and the routing potential}

Both stages of the record are guided by a potential over the pairs that
remain to be covered,
\begin{equation}
  \Phi(\pi) \;=\; \sum_{(a,c)\ \text{uncovered}} w\big(d(\pi(a),\pi(c))\big),
  \qquad w(0)=w(1)=0,
  \label{eq:supp-layout-potential}
\end{equation}
where $d$ is the distance on the grid.

\paragraph{Placement.} The initial mapping is obtained by simulated
annealing over bijections, minimising Eq.~\eqref{eq:supp-layout-potential}
over all pairs with the linear potential $w(d)=d-1$, so that adjacent pairs
are free and every other pair is charged its excess distance. A move
exchanges the qubits on two sites, either neighbouring or chosen at random,
and the temperature decreases geometrically. Several annealed placements are
ranked by the length of a short routing run from each, and the best two are
retained. The agents found that annealing the placement on the fitted
potential described below, rather than on the linear one, gave longer
routes.

\paragraph{Greedy routing.} From a mapping $\pi$, every grid edge $e$ is a
candidate SWAP and is assigned the score
\begin{equation}
  s(e) \;=\; A\, f(e) \;+\; \big[\Phi(\pi)-\Phi(\sigma_e\circ\pi)\big],
  \label{eq:supp-layout-score}
\end{equation}
where $f(e)\le 6$ is the number of pairs the SWAP covers and $A$ is a bonus
per covered pair. The greedy walk applies the highest-scoring SWAP, applies
every term whose qubits have become adjacent, and repeats until all pairs
are covered. The potential is maintained incrementally as an $N\times N$
table of partial sums, so that the scores of all grid edges are obtained in
one vectorised operation.

\paragraph{The shape of the potential.} The first submission used
$w(d)=d-1$, and its agent reported that convex powers $(d-1)^\beta$ with
$\beta>1$ are clearly worse and that exponents between $0.5$ and $1$ make
little difference. The record replaced hand-set shapes with a fitted one.
With the normalisation $w(2)=1$, which removes the redundant overall scale
of Eq.~\eqref{eq:supp-layout-score}, the eight increments of $w(3),\dots,
w(10)$ and the bonus $A$ were fitted by an evolution
strategy~\citep{beyer2002evolution}. The objective was the mean route length
on training instances that are not part of the leaderboard ($8\times8$
graphs with other random seeds), evaluated first with the greedy walk and
then refined with a beam search. Separate fits were obtained for
$p=0.25,0.5,0.75$, and the shape used for an instance is interpolated in
its density $m/\binom{N}{2}$.

The fitted potentials rise steeply up to distance three and flatten beyond
it (Fig.~\ref{fig:supp-layout-analysis}a). Bringing a pair from distance
three to distance two lowers the potential by $1.0$ to $1.5$, at least as much
as the final step to adjacency, which lowers it by $1$ and earns the bonus
$A$ of $0.7$ to $1.0$. Pairs that are far apart attract each other only
weakly, by $0.25$ to $0.4$ per unit of distance on average beyond distance
five. The fitted potential alone shortens
the greedy walk by $6\%$ on the $8\times8$ instances
(Fig.~\ref{fig:supp-layout-analysis}b). The independent reimplementation
arrived at a similar $w$ but a much smaller bonus, $A\approx0.25$. On its
training instances, and with the same beam search as evaluator, its shape
shortened the routes by $10\%$ relative to the linear potential and the
record's shape by $7\%$. That two different balances between the bonus and
the pull of the potential perform comparably suggests that the route length
depends only weakly on this balance.

\subsubsection{Rollout-ranked beam search}

\begin{figure}[ht]
  \centering
  \includegraphics[width=0.74\textwidth]{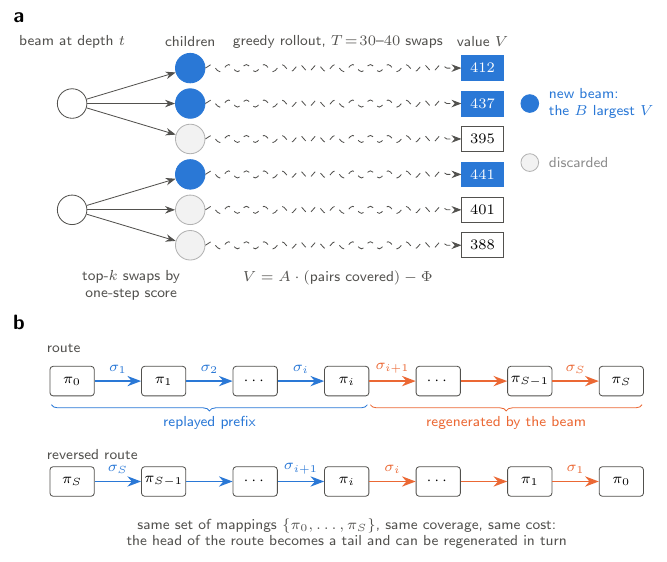}
  \caption{\textbf{The two search operations of the record.}
  \textbf{a},~Rollout-ranked beam search. Each state in the beam proposes its
  $k$ best SWAPs under the one-step score. Each child is advanced by
  $T=30$--$40$ steps of the greedy walk, and the $B$ children with the
  largest value $V$ (Eq.~\ref{eq:supp-layout-value}) at the end of the rollout
  form the next beam (blue). The numbers are illustrative.
  \textbf{b},~Local search in both time directions. A walk is replayed to a
  random step $i$ and the remainder is regenerated by a narrow beam search;
  the new walk is kept if it is shorter. Because each SWAP is its own
  inverse, the reversed sequence of SWAPs, started from the final mapping
  $\pi_S$, visits the same mappings in reverse order and therefore covers
  the same pairs at the same cost. Reversing the walk turns its beginning
  into an end that can be regenerated in the same way.}
  \label{fig:supp-layout-search}
\end{figure}

A beam search over walks keeps the $B$ best partial walks and extends each by
its $k$ highest-scoring SWAPs. In the first submission the children were
ranked by the cumulative score of Eq.~\eqref{eq:supp-layout-score}, a
one-step criterion that favours SWAPs which cover pairs immediately. The
record ranks each child instead by the value
\begin{equation}
  V \;=\; A\cdot(\text{pairs covered}) \;-\; \Phi
  \label{eq:supp-layout-value}
\end{equation}
of the state reached after $T=30$ to $40$ further steps of the greedy walk
(Fig.~\ref{fig:supp-layout-search}a), with an additional credit for walks
that finish within the $T$ steps. This is a rollout in the sense of
approximate dynamic programming~\citep{bertsekas1997rollout}: a fixed base
policy, the greedy walk, is used to estimate the value of each candidate, and
the search prefers states from which the greedy walk continues to cover
pairs efficiently over states that cover pairs immediately. Mappings that
have already been visited are discarded, and the walk returns as soon as a
child covers the last pair. The beam is narrow, with $B=6$ to $8$ and
$k=2$ or $3$.

The rollouts make each beam search an order of magnitude more expensive, and
the time budget of $18$ seconds per instance at $8\times8$ forces a choice
between one rollout-ranked search and many inexpensive ones. The agent
measured that, within this budget, a rollout-ranked beam alone performs
about as well as the first submission's strategy of running many
diversified plain beams and keeping the best; the final program therefore
runs both, in that order, followed by local search.

\subsubsection{Local search in both time directions}

The best walk found is improved by regenerating its end. The walk is
replayed up to a random step $i$, the pairs not yet covered are recomputed,
and the remaining $S-i$ SWAPs are replaced by a narrow, randomised beam
search that is accepted only if it is shorter. This operation can only
modify the end of a walk. The agents observed that it applies equally to the
beginning, because coverage depends only on the set of mappings that a walk
visits: since $\sigma_t^{-1}=\sigma_t$, the walk
$\pi_S,\pi_{S-1},\dots,\pi_0$ produced by applying $\sigma_S,\dots,\sigma_1$
to the final mapping $\pi_S$ is a valid solution with the same cost, whose
initial mapping is $\pi_S$. Reversing the walk with probability one half
before each regeneration therefore lets the same operation act on either
end (Fig.~\ref{fig:supp-layout-search}b), and in particular revises the
initial placement in the light of the route that follows it.

A final pruning pass removes any single SWAP, and any pair of SWAPs on the
same grid edge, whose deletion leaves every pair covered; each deletion is
checked by replaying the walk.

\subsubsection{What the measurements show}

\begin{figure}[ht]
  \centering
  \includegraphics[width=\textwidth]{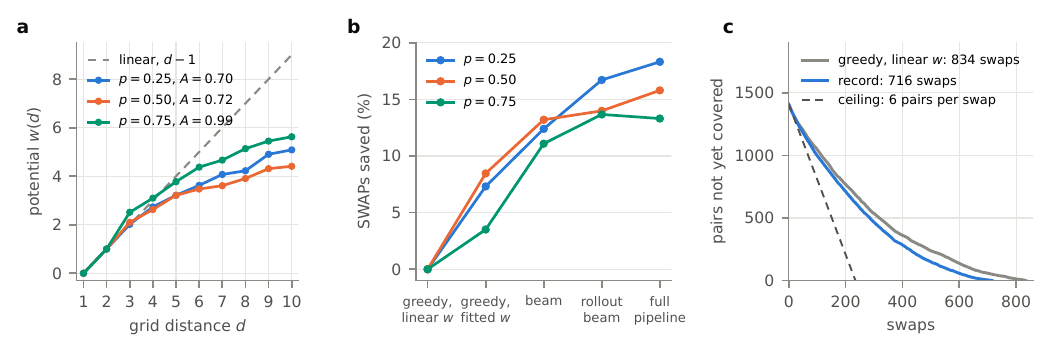}
  \caption{\textbf{Anatomy of the record at $8\times8$.} \textbf{a},~Routing
  potentials $w(d)$ fitted by an evolution strategy for the three densities,
  with the corresponding bonus $A$, compared with the linear potential
  $d-1$. \textbf{b},~Reduction in the number of SWAPs relative to the greedy
  walk with the linear potential, for increasingly complete versions of the
  method, averaged over the three instances at each density. All routes
  except the last start from the same annealed placement; the complete
  program anneals its own. \textbf{c},~Pairs not yet covered along the walk
  for the instance \texttt{er-8x8-p75-s0} ($m=1516$), for the greedy walk with
  the linear potential and for the record, compared with the maximum rate of
  six pairs per SWAP that underlies Eq.~\eqref{eq:supp-layout-bound}. All
  routes were verified by the frozen evaluator.}
  \label{fig:supp-layout-analysis}
\end{figure}

\paragraph{Contribution of each component.} To separate the contributions,
we fixed one annealed placement for each of the nine $8\times8$ instances and
routed from it with increasingly complete versions of the method
(Fig.~\ref{fig:supp-layout-analysis}b). Relative to the greedy walk with the
linear potential and $A=1$, the fitted potential reduces the mean number of
SWAPs by $6.1\%$, a beam search with $B=16$ and $k=4$ by $12.1\%$, the
rollout-ranked beam by $14.5\%$, and the complete program, which also
anneals its own placements and applies local search and pruning, by
$15.4\%$. Because the search is limited by wall-clock time, repeated runs
of the complete program differ by about $2\%$ per instance, and the mean of
$560.6$ obtained here is close to the registered $552.4$.

\paragraph{The cost of a walk lies at its end.} The rate at which a walk
covers pairs falls steadily along it (Fig.~\ref{fig:supp-layout-analysis}c).
On an $8\times8$ instance with $p=0.75$ and $m=1516$, the record covers $4.3$
pairs per SWAP in the first tenth of its walk and $0.4$ in the last tenth.
Ninety percent of the pairs are covered after $60\%$ of the SWAPs, and $43\%$
of all SWAPs cover at most one pair. The bound of
Eq.~\eqref{eq:supp-layout-bound} assumes six pairs per SWAP throughout and is
correspondingly far from the walks found: $234$ against $716$ on this
instance. The remaining pairs at the end of a walk are scattered over the
grid at distances three to seven, and covering them dominates the cost.

\paragraph{Negative results.} The agents' notes record several approaches
that were measured and rejected, each with a quantitative reason.
\begin{itemize}
  \item \emph{Re-embedding the remaining graph.} An appealing strategy for
  the end of a walk is to anneal a new placement for the pairs that remain,
  route to it, and cover the embedded pairs at no further cost. On an
  $8\times8$ instance with $p=0.75$, $141$ pairs remain after $70\%$ of the
  walk; at most $74$ of them can be embedded in the grid simultaneously, and
  reaching such an embedding requires a total displacement of the qubits of
  about $324$ grid steps, more than the $227$ SWAPs with which the greedy
  walk completes the route. Random interaction graphs leave no stage of the
  walk at which this strategy pays.
  \item \emph{Learned cost-to-go.} A regression of the remaining number of
  SWAPs on features of the state fits with $R^2=0.99$ once the number of
  qubits with an uncovered partner is included. Adding this feature to the
  score as a per-qubit bonus has no effect: it measures progress but does
  not indicate which SWAP to take.
  \item \emph{Structured sweeps.} Oblivious odd--even reversals of whole rows
  or columns, as in the swap network, cover fewer pairs per SWAP than the
  adaptive greedy walk at every stage of the route.
  \item \emph{Compute.} A tenfold increase in beam width or rollout length
  reduces the SWAP count by about $6\%$ at $p=0.75$ and $2$ to $3\%$ at
  $p=0.5$. The independent reimplementation made its router $11\%$ faster
  and observed no change in the result. Further progress within this family
  of methods requires a new idea rather than more search.
\end{itemize}

\paragraph{Calibration and limitations.} On the three $3\times3$ instances
for which the exhaustive solver built with the challenge gives the optimum ($1$, $1$ and $2$
SWAPs), the record attains it; the independent reimplementation attains the
optimum on every $2\times2$ and $2\times3$ instance tested. On larger grids
no optimum is known, and the ratio between the record and the bound of
Eq.~\eqref{eq:supp-layout-bound} grows from $2.6$ at $3\times3$ to $3.7$ at
$8\times8$; it is not known how much of this gap is due to the bound. The
circuit depth was not optimised. At $8\times8$ the record has a mean depth
of $377$, against $129$ for the swap network, whose SWAPs are executed in
parallel by construction.

\subsection{Unitary synthesis}
\label{supp:sol-unitary}

One agent submitted to this challenge, and its submission holds the record on
all fourteen instances, $d2$--$d8$ (diagonal, $A=I$) and $p2$--$p8$ (a fixed
invertible map $A\neq I$). The submission consists of one verified circuit per instance,
together with a description of the method that produced them; the search
program itself was not submitted. Its value therefore lies less in the
program than in two ideas, which we state below as exact results, and in the
circuits, which we examine with an exhaustive search of our own. Coefficients
of phase polynomials are written in units of $\pi/4$, so that a target is a
function $c$ from nonempty subsets $S\subseteq\{0,\dots,n-1\}$ to
$\mathbb{Z}_8$, with phase $\phi(x)=\tfrac{\pi}{4}\sum_S c_S\,\chi_S(x)$.

\subsubsection{Circuits as walks through bases}

During a circuit over $\{\mathrm{CNOT},R_z\}$ each wire carries a parity of
the input bits: if wire $q$ carries the row vector $r_q\in\mathbb{F}_2^n$, its
computational-basis value is $r_q\cdot x$. Initially $r_q=e_q$. A CNOT with
control $c$ and target $t$ replaces $r_t$ by $r_t\oplus r_c$, and an
$R_z(\theta)$ on wire $q$ adds $\theta\,\chi_{r_q}(x)$ to the phase. The
ordered rows $(r_0,\dots,r_{n-1})$ form an invertible matrix, and a circuit
with $S$ CNOTs is therefore a walk of length $S$ from $I$ to the final matrix
in the Cayley graph of $\mathrm{GL}(n,\mathbb{F}_2)$ generated by the
$n(n-1)$ elementary row additions. The circuit implements
$|x\rangle\mapsto e^{i\phi(x)}|Ax\rangle$ exactly when the walk ends at $A$
and $\phi$ is a combination of the parities that the walk visits, with any
angles~\citep{amy2018controlled}. Since $R_z$ gates do not enter the
objective, the CNOT count is the length of the walk.

\begin{figure}[ht]
  \centering
  \includegraphics[width=0.86\textwidth]{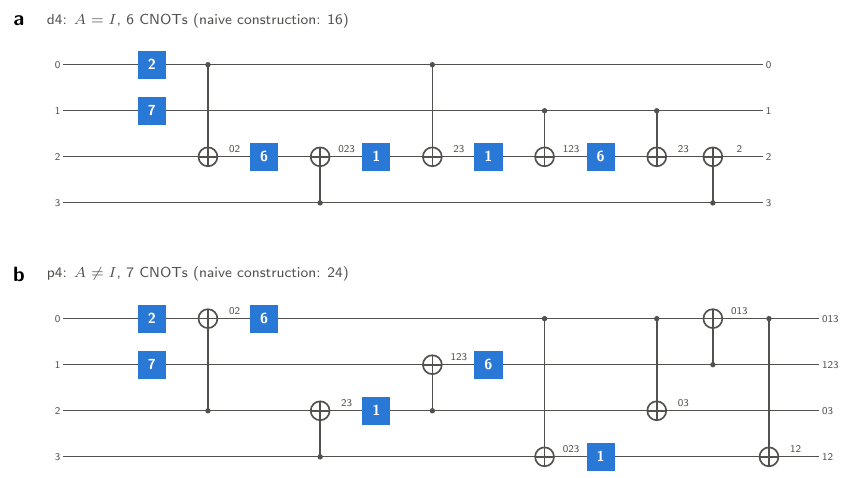}
  \caption{\textbf{Circuits of the record for $n=4$.} Wires are labelled at
  the left by the bit they initially carry and at the right by the parity they
  carry at the end; after each CNOT, the new parity of its target wire is
  shown above the wire (for example, $023$ denotes $x_0\oplus x_2\oplus
  x_3$). Each $R_z$ is applied the first time its parity appears, and its
  angle is given in units of $\pi/4$. The coefficients are those of the
  rewritten polynomial of Fig.~\ref{fig:supp-unitary-rewrite}b.
  \textbf{a},~Diagonal target: wire $2$ walks through
  $2\to02\to023\to23\to123\to23\to2$ and the walk closes at $I$.
  \textbf{b},~The same phase polynomial composed with the map $A$ of
  instance $p4$. The walk ends at $A$; realising $A$ alone requires six CNOTs,
  so the phase polynomial costs one additional CNOT. Both circuits are
  optimal (see text).}
  \label{fig:supp-unitary-walks}
\end{figure}

The reference construction treats the two requirements separately: it
computes and uncomputes each parity of $\phi$ with its own ladder of CNOTs,
returning to $I$ each time, and then synthesises $A$. The submission instead
searches for one walk that visits every required parity and ends at $A$
(Fig.~\ref{fig:supp-unitary-walks}). This is the same structure as in the
layout challenge, where a solution is a walk through the symmetric group,
generated by SWAPs along grid edges, that must make every interacting pair
adjacent at some step; here the walk runs through $\mathrm{GL}(n,\mathbb{F}_2)$
and must make every required parity appear on some wire.

\subsubsection{Equivalent phase polynomials}

Because the angles are multiples of $\pi/4$, many phase polynomials define the
same unitary, and the set of parities that a walk must visit depends on which
representative is chosen. The submission characterises the equivalent
representatives by their \emph{containment moments}
$M_T(c)=\sum_{S\supseteq T}c_S$.

\paragraph{Proposition.} Two phase polynomials $c$ and $c'$ with integer
coefficients define the same unitary if and only if
\begin{equation}
  M_T(c)\equiv M_T(c') \pmod{2^{\,4-|T|}}\qquad\text{for every } T
  \text{ with } 1\le|T|\le3 ,
  \label{eq:supp-unitary-moments}
\end{equation}
that is, singleton moments agree modulo $8$, pair moments modulo $4$ and
triple moments modulo $2$.

\emph{Proof.} Iterating $a\oplus b=a+b-2ab$ gives, for $x\in\{0,1\}^n$,
\begin{equation}
  \chi_S(x)=\sum_{\emptyset\neq T\subseteq S}(-2)^{|T|-1}\prod_{k\in T}x_k ,
  \qquad\text{hence}\qquad
  \sum_S d_S\,\chi_S(x)=\sum_{T\neq\emptyset}(-2)^{|T|-1}M_T(d)\prod_{k\in T}x_k .
\end{equation}
Both unitaries apply phase $0$ to $x=0$, so they coincide exactly when
$d=c-c'$ satisfies $\sum_S d_S\chi_S(x)\equiv0\pmod 8$ for every $x$. A
function on $\{0,1\}^n$ has a unique multilinear expansion, whose
coefficients are integer combinations of its values (Möbius inversion), so
this holds if and only if every coefficient $(-2)^{|T|-1}M_T(d)$ is divisible
by $8$. For $|T|\ge4$ this is automatic, and for $|T|=1,2,3$ it is
Eq.~\eqref{eq:supp-unitary-moments}.\hfill$\square$

\paragraph{Elementary rewrites.} Inverting the moment map shows that the
polynomials equivalent to zero are generated, modulo $8$, by one identity for
each subset $T$ with $|T|\ge2$:
\begin{align}
  4\chi_{ab} &\equiv 4\chi_a+4\chi_b, \nonumber\\
  2\chi_{abc} &\equiv 2(\chi_{ab}+\chi_{ac}+\chi_{bc})-2(\chi_a+\chi_b+\chi_c),
  \label{eq:supp-unitary-rewrites}\\
  0 &\equiv \textstyle\sum_{\emptyset\neq S\subseteq T}(-1)^{|S|}\chi_S
  \qquad(|T|\ge4). \nonumber
\end{align}
The first identity removes a CNOT-requiring term of angle $\pi$ entirely,
since single-qubit parities are present on the wires at the start; it is the
example given in the main text. The second and third trade one parity for
several others and are useful when those others lie on a short walk.
Counting the moduli in Eq.~\eqref{eq:supp-unitary-moments} shows that each
target has
$2^{\,3(2^n-1)-3n-2\binom n2-\binom n3}$ equivalent representatives modulo $8$:
$2^{17}$ for $n=4$ and $2^{629}$ for $n=8$. The same freedom underlies T-count
optimisation, where it is used to minimise the number of odd
coefficients~\citep{amy2019tcount}; here it is used to choose a support that
a short walk can visit.

The same argument with real coefficients shows that the moments of any
difference must satisfy $M_T(d)\in2^{\,4-|T|}\mathbb{Z}$ for every $T$. For
$n\le4$ this forces all coefficients of $d$ to be integers, so $R_z$ angles
that are not multiples of $\pi/4$ cannot reduce the CNOT count; for $n\ge5$,
finer angles on parities of weight at least five are admissible in
principle.

\begin{figure}[ht]
  \centering
  \includegraphics[width=0.9\textwidth]{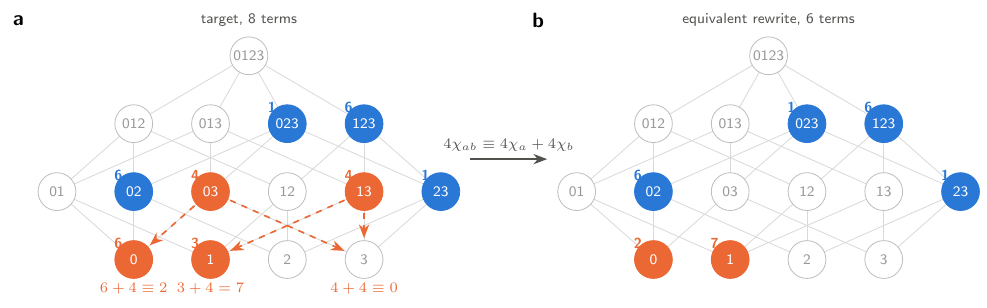}
  \caption{\textbf{Rewriting the phase polynomial of instance $d4$.} Nodes
  are the nonempty subsets of $\{0,1,2,3\}$, ordered by inclusion; filled
  nodes carry a nonzero coefficient, shown in units of $\pi/4$.
  \textbf{a},~The target has eight terms, two of which are pair terms with
  coefficient $4$. By the first identity of
  Eq.~\eqref{eq:supp-unitary-rewrites}, each is replaced by coefficient $4$
  on its two qubits (dashed). The two contributions to qubit $3$ cancel
  modulo $8$. \textbf{b},~The rewritten polynomial has six terms and only
  four parities that require CNOTs, which the walk of
  Fig.~\ref{fig:supp-unitary-walks}a visits with six CNOTs. Changed
  coefficients are shown in orange.}
  \label{fig:supp-unitary-rewrite}
\end{figure}

\subsubsection{The search and its results}

According to the submission's notes, small instances were solved exactly by
a bounded search over walks, and larger ones by randomised routing, local
resynthesis of segments of a walk, and exact searches on short windows, while
the search ranged over equivalent representatives. Every stored circuit was
checked on all $2^n$ basis states before submission, and again by the frozen
evaluator. Table~\ref{tab:supp-unitary} compares the circuits with the naive
construction and with a stronger baseline built during the construction of
the challenge, which reuses one ladder for terms whose supports are nested.

\begin{table}[h]
  \centering\small
  \caption{\textbf{CNOT counts for unitary synthesis.} ``Terms'' is the number
  of terms of the target phase polynomial; ``$R_z$'' is the number of terms of
  the representative used by the record, which equals its number of $R_z$
  gates. ``Nested'' reuses a ladder for terms whose supports are nested; ``Opt.''
  is the minimum over all circuits over $\{\mathrm{CNOT},R_z\}$ with arbitrary
  angles, computed by exhaustive search for $n\le4$.}
  \label{tab:supp-unitary}
  \setlength{\tabcolsep}{4.5pt}
  \begin{tabular}{cc|ccccc|ccccc}
    \toprule
    & & \multicolumn{5}{c|}{Diagonal, $A=I$} & \multicolumn{5}{c}{Non-diagonal, $A\neq I$} \\
    $n$ & Terms & $R_z$ & Naive & Nested & Record & Opt. & $R_z$ & Naive & Nested & Record & Opt. \\
    \midrule
    2 & 3  & 3  & 2  & 2  & \textbf{2}  & 2  & 3  & 4  & 4  & \textbf{2}  & 2 \\
    3 & 6  & 6  & 10 & 8  & \textbf{4}  & 4  & 6  & 11 & 9  & \textbf{3}  & 3 \\
    4 & 8  & 6  & 16 & 12 & \textbf{6}  & 6  & 6  & 24 & 20 & \textbf{7}  & 7 \\
    5 & 10 & 5  & 26 & 20 & \textbf{10} & -- & 5  & 37 & 31 & \textbf{10} & -- \\
    6 & 12 & 8  & 26 & 22 & \textbf{13} & -- & 8  & 44 & 40 & \textbf{16} & -- \\
    7 & 14 & 13 & 46 & 40 & \textbf{19} & -- & 11 & 67 & 61 & \textbf{24} & -- \\
    8 & 16 & 13 & 46 & 42 & \textbf{23} & -- & 20 & 73 & 69 & \textbf{25} & -- \\
    \bottomrule
  \end{tabular}
\end{table}

\paragraph{The linear map is nearly free.} In the naive construction the map
$A$ is synthesised after the phase polynomial, at a cost of $27$ CNOTs at
$n=8$; in the record it costs $2$ ($25$ against $23$). At $n=3$ the
non-diagonal target is even cheaper than the diagonal one, $3$ CNOTs against
$4$, because a diagonal target requires a closed walk, whereas ending at
$A\neq I$ removes the obligation to return. At $n=4$ the map alone requires
six CNOTs, the distance from $I$ to $A$ in the Cayley graph, and the complete
target seven.

\paragraph{The representative is chosen for the walk, not for sparsity.} The
rewriting reduces the number of terms for every $n\ge4$ except at $p8$, for
example from $10$ to $5$ at $n=5$. At $p8$, however, the chosen representative
has $20$ terms against $16$ in the target: the search accepted more $R_z$ gates in exchange for parities that lie
on a shorter walk. The $R_z$ count is reported by the evaluator but was not
optimised.

\paragraph{Optimality for $n\le4$.} We determined the minimum CNOT count
over all circuits over $\{\mathrm{CNOT},R_z\}$ for $n\le4$. For each target we
enumerated its equivalent representatives modulo $8$ ($2^{17}$ at $n=4$) and
recorded the minimal sets of parities that some representative is supported
on; by the preceding paragraph, arbitrary angles reduce to this case. A
breadth-first search over pairs (current basis, set of visited parities) then
finds the shortest walk that ends at $A$ and whose visited set contains one of
these minimal supports. A walk is discarded when its distance to $A$ in the
Cayley graph exceeds the remaining length, which does not affect the result.
The record attains the minimum on all six instances with $n\le4$
(Table~\ref{tab:supp-unitary}). For $n\ge5$, where
$|\mathrm{GL}(5,\mathbb{F}_2)|\approx10^7$, the optimum is not known.

\paragraph{Limitations.} The submission stores one circuit for each of the
fourteen fixed targets. It establishes that these circuits exist and are
correct, and the two ideas above explain why they are short, but it does not
demonstrate a general algorithm: the program that found them was not
submitted and cannot be run on new targets. A version of the challenge with
held-out targets would measure the method rather than the constructions.

\end{document}